%% file: Black-Holes-with-Halos-submit.tex
\documentclass[11pt,oneside,letterpaper]{article}
\usepackage{amssymb}
\usepackage{amsmath}
\usepackage[dvips]{graphicx}
\usepackage{setspace}
\usepackage{fancyhdr}
\usepackage{xcolor}
\usepackage{ifpdf}
\usepackage{graphicx}
\usepackage{rotating}
\usepackage{comment}
\usepackage{braket}
\usepackage[pdfusetitle,bookmarks,pdfpagelabels,breaklinks,plainpages=false,pdfpagemode=UseNone]{hyperref}
\usepackage{cleveref}
 \usepackage{fixltx2e}
 \usepackage[utf8x]{inputenc}

\usepackage{datetime,color}
\usepackage{graphicx}
\usepackage{float}
\usepackage{authblk}
\usepackage[bf,hang,footnotesize]{caption}
\usepackage[title]{appendix}

\newcommand{\be}{\begin{equation}}

\newcommand{\ee}{\end{equation}}
\newcommand{\re}{\ensuremath{{\text{Re}}}}
\newcommand{\im}{\ensuremath{{\text{Im}}}}

\newcommand{\order}[1]{{\ensuremath{\mathcal{O} \left( #1 \right)}}}
\newcommand{\eq}{{\ensuremath{{}={}}}}
\newcommand{\pheq}{{\ensuremath{\hphantom{{}={}}}}}
\newcommand{\td}{\ensuremath{{\text{d}}}}
\newcommand{\cN}{\ensuremath{{\mathcal{N}}}}

\title{\texorpdfstring{\vspace{60pt}}{}\textsc{Black Holes with Halos} \texorpdfstring{\vspace{30pt}}{}}
\author[$\sharp$,$\dag$]{Ruben Monten}
\author[$\dag$]{Chiara Toldo \vspace{10pt}}
\affil[$\dag$]{\it \footnotesize Department of Physics, Columbia University, 538 West 120th Street, New York, New York 10027}
\affil[$\sharp$]{\it \footnotesize Institute for Theoretical Physics, KU Leuven, 3001 Leuven, Belgium}
\date{}

\begin{document}
  \maketitle
  \vspace{60pt}

  \begin{abstract}
  
  \vspace{5mm}
  
    \noindent
    We present new AdS$_4$ black hole solutions in $\mathcal{N} =2$ gauged supergravity coupled to vector and hypermultiplets. We focus on a particular consistent truncation of M-theory on the homogeneous Sasaki-Einstein seven-manifold $M^{111}$, characterized by the presence of one Betti vector multiplet. We numerically construct static and spherically symmetric black holes with electric and magnetic charges, corresponding to M2 and M5 branes wrapping non-contractible cycles of the internal manifold. These configurations have nonzero temperature and are moreover surrounded by a massive vector field halo. For these solutions we verify the first law of black hole mechanics and we analyze the thermodynamics and phase transitions in the canonical ensemble, interpreting the process in the corresponding dual field theory.

  \end{abstract}
  \newpage
  
  \tableofcontents
    
  \section{Introduction}
    \input{Introduction2}

  \section{Setup}
    \input{Setup}

  \section{Finding black hole solutions}
    \input{Solutions}

  \section{Black hole thermodynamics}
    \input{FirstLaw}

  \section{Canonical ensemble}
    \input{PhaseTransitions}

  \section{Outlook}
    \input{Outlook}

  \section*{Acknowledgments}
    We would like to thank D. Anninos, A. Bzowski, R. Emparan, A. Gnecchi, N. Halmagyi, I. Klebanov, Y. Korovin, I. Papadimitriou, M. Rangamani, S. Vandoren, O. Varela,  A. Zaffaroni,  for very useful discussions and correspondence, and in particular F. Denef for important insights throughout the development of this project. The work of RM is supported by FWO Flanders. CT acknowledges support from the NWO Rubicon Grant, Columbia University and from DOE grant DE-SC0011941.
  
  \begin{appendices}
  \addtocontents{toc}{\protect\setcounter{tocdepth}{0}}
    \section{Further notations and conventions}
      \label{sec:conventions}
      \input{Conventions}

    \section{Equations of motion}
      \label{sec:EoM}
      \input{EoM}
  \end{appendices}

  \bibliographystyle{myJHEP}
  \bibliography{tmp}

\end{document}

%% file: Introduction2.tex
The analysis of Anti-de Sitter (AdS) black hole solutions in theories of four-di\-men\-sio\-nal gauged supergravity is important for at least two reasons. On one hand, the AdS/CFT correspondence sheds light on the microstate structure of the supersymmetric configurations. In this regard, some recent developments \cite{Benini:2015eyy,Benini:2016rke} successfully matched the BPS black hole \cite{Cacciatori:2009iz} entropy with the ground state degeneracy of the corresponding twisted ABJM \cite{Aharony:2008ug} theory, via supersymmetric localization. On the other hand, AdS black holes from string theory provide interesting gravitational backgrounds for top-down holographic approaches: one can map the rich thermodynamics and phase transitions of these systems to processes in the dual field theory, providing a description of strongly coupled field theoretical phenomena, such as superconductivity \cite{Gubser:2008px,Hartnoll:2008vx,Hartnoll:2008kx}. 

The characterization and construction of solutions of gauged supergravity models coming from M-string theory is an important step in this direction. So far, much of the effort has been directed towards the analysis and characterization of black hole solutions of $\mathcal{N}=2$ Abelian Fayet-Iliopoulos gauged supergravity \cite{Dall'Agata:2010gj,Hristov:2010ri,Kachru:2011ps,Donos:2011pn,Meessen:2012sr,Hristov:2012nu,Hristov:2013spa,Halmagyi:2013qoa,Gnecchi:2013mta,Barisch-Dick:2013xga,Klemm:2015xda,Katmadas:2015ima}. The first example of static supersymmetric AdS$_4$ black holes was analytically constructed in \cite{Cacciatori:2009iz}, while previous studies \cite{Duff:1999gh} yielded naked singularities. In this model, the scalars are uncharged under the gauge group and solution-generating techniques of ungauged supergravity can be used to construct new configurations (see for instance \cite{Klemm:2012yg,Klemm:2012vm,Gnecchi:2012kb,Halmagyi:2013uza}). 

The construction of analytic black hole solutions in other models of gauged supergravity, in particular those including hypermultiplets, initiated in \cite{Meessen:2012sr,Chimento:2015rra,Klemm:2016wng}, revealed to be much harder since the matter content includes charged scalars and massive vectors. Charged scalars and massive vectors are a generic feature of AdS$_4 \times M^6$ compactifications dual to ABJM theory. In these models a linear combination of the U(1) gauge fields obtained from the reduction of the RR fields becomes massive due to the Higgs mechanism. This was shown in the original example for the compactification on $CP^3$ \cite{Aharony:2008ug}, and the Higgsing also occurs in other models (see for instance \cite{Gauntlett:2009zw,Cassani:2012pj}) arising from compactifcations of M-theory on 7d Sasaki-Einstein manifold.

The fact that a $U(1)$ is Higgsed has nontrivial consequences for black hole physics, and in particular for the analysis of black hole bound states in AdS$_4$ \cite{Anninos:2013mfa}. A prerequisite for the existence of multi-centered black holes is that the electromagnetic interaction balances the gravitational one. A massive vector field decays exponentially, rather than polynomially, and this generally modifies the conditions for a bound state to exist. Moreover, bound configurations with magnetic charges would come with strings attached \cite{Anninos:2013mfa} due to the Meissner effect. All these ingredients can in principle play an important role in the existence and stability of these bound states. 

\vspace{3mm}

The aim of this paper is take the first steps to address these problems, by constructing  AdS$_4$ thermal black holes with an embedding in M-theory, surrounded by massive vectors and charged scalars. These solutions will provide suitable thermal backgrounds for the subsequent study of the probe stability.

We focus our attention on reductions of eleven dimensional supergravity whose vacua preserve $\mathcal{N} =2$ supersymmetry. Such consistent truncations of M-theory on homogeneous seven-dimensional Sasaki--Einstein manifolds with $SU(3)$ structure were found in \cite{Cassani:2012pj}. We work with a specific reduction of 11d supergravity on the $SE_7$ manifold $M^{111}$, with field theory dual in the class of \cite{Jafferis:2009th,Benini:2009qs}. This truncation has a massive vector in its spectrum, which corresponds to a broken global symmetry in the dual field theory. Furthermore, it is characterized by the presence one Betti vector multiplet, dual to a global baryonic symmetry. On the gravity side, this multiplet contains light degrees of freedom, in particular massless vectors and scalars with mass $m^2 l^2= -2$. 

Zero-temperature, 1/4 BPS black hole solutions for various models, including $Q^{111}$ and $M^{111}$, were found in \cite{Halmagyi:2013sla}, in the form of  flows from $AdS_4$ to $AdS_2 \times S^2$ near-horizon geometries, by solving the BPS equations. Solutions of the same models, with planar horizons were previously obtained in \cite{Herzog:2009gd,Klebanov:2010tj,Donos:2012sy}. The presence of charged scalars considerably complicates the equations, hence the flows were obtained mostly numerically. 

The black holes we present here correspond to nonzero temperature generalizations of the black holes of \cite{Halmagyi:2013sla} and are found by solving the Einstein, Maxwell and scalar equations of motion. This reduces to a boundary value problem for a system of 14 coupled ODEs, which we solve numerically using a shooting method.

\begin{figure}[htbp]
\centering
{\includegraphics[width=80mm]{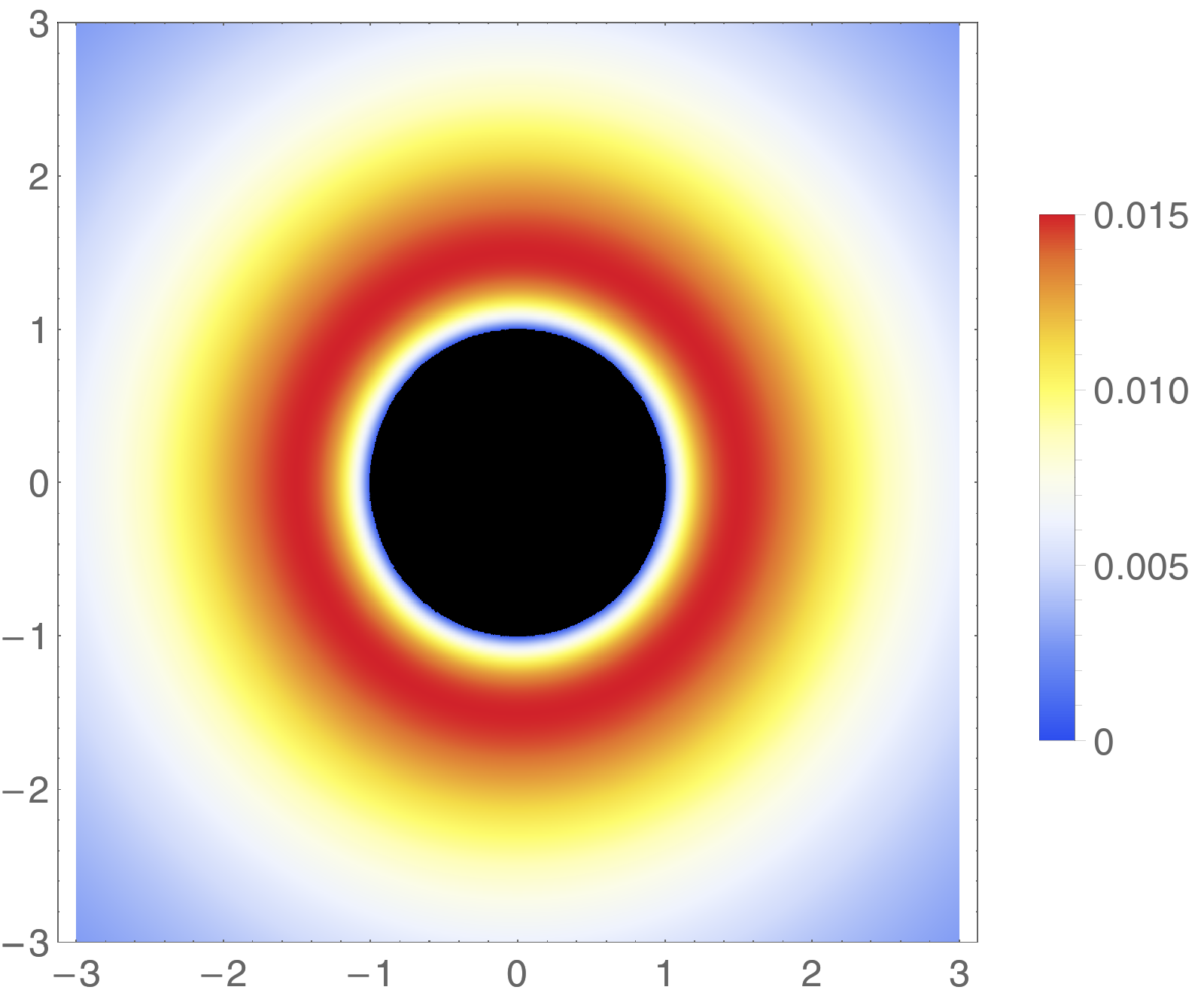}}
\caption{2D plot of the massive vector field profile for a electric solution (details of the configuration are provided in Figure \ref{figure_electric}, Section \ref{sec:examples}). The massive vector profile is peaked outside the black hole, forming a ``halo'' or atmosphere surrounding the black hole. \label{Fig2d}}
\end{figure}
We were able to construct dyonic AdS$_4$ black hole solutions with nontrivial matter profiles outside the horizon\footnote{The no-hair theorem of Bekenstein \cite{Bekenstein:1971hc} rules out massive vector field hair in four-di\-men\-sio\-nal asymptotically flat space-time. However, interactions among the different fields and AdS asymptotics are sufficient to evade the theorem. For further work on black holes and branes with massive vector fields, but Lifshitz asymptotics see \cite{Korovin:2013nha}. Work on AdS black holes with massive vectors in $d>4$ can be found in \cite{Liu:2014tra}.}. In particular, a massive vector field ``halo'' surrounds the black hole solutions, as depicted in Figure \ref{Fig2d}. The solutions asymptotically approach the AdS vacuum in which the vector remains massive, in contrast to the case of holographic superconductors \cite{Gubser:2008px,Hartnoll:2008vx,Hartnoll:2008kx}, where a $U(1)$ symmetry is broken in the proximity of the horizon and is restored in the UV. We find that the presence of Betti vector multiplets is required in order to find (non-extremal) black hole solutions with nontrivial behavior of the massive vector. In the M-theory picture, these additional electric and magnetic charges correspond to wrapped M2 and M5-branes on cycles of the internal manifold. 

To verify the accuracy of our numerics, we have checked that the first law of thermodynamics is satisfied on our solutions. We have performed holographic renormalization to compute the renormalized on shell action and subsequently studied the behavior of the free energy and its non-analytic points, searching for phase transitions. 

\vspace{2mm}

The analysis of the stability of charged probe black holes in the background of these new configurations, along the lines of \cite{Anninos:2011vn,Anninos:2012gk,Chowdhury:2013ema,Anninos:2013mfa}, in view of the possible description of the holographic vitrification process is work in progress, and will be presented in a forthcoming paper. More directions in this regard will be presented in the outlook section.

%% file: Setup.tex
\subsection{Model \texorpdfstring{$M^{111}$}{M111}}

The setup for our computations is the Abelian four-dimensional $\mathcal{N}=2$ gauged supergravity theory obtained upon reduction of eleven-dimensional  supergravity on the 7d Sasaki-Einstein manifold $M^{111}$. This is the coset manifold $G/H$ where $G= SU(3) \times SU(2)$ and $H= SU(2) \times U(1)$. Its second Betti number is $b_2(M^{111}) = 1$, hence there is one nontrivial two-cycle around which $M_2$ branes can wrap. The effective field theory therefore contains one Betti vector multiplet in its spectrum. The same truncation can alternatively be obtained from the reduction on the seven-dimensional manifold $Q^{111}$ (with $G=SU(2)^3$ and $H= U(1)^2$, and $b_{2}(Q^{111})=2$ hence two Betti multiplets), provided we consistently truncate one of the two Betti multiplets by suitably identifying two vectors and two scalar fields. The superconformal field theory dual to the $Q^{111}$ model is the superconformal Chern-Simons flavored quiver of \cite{Jafferis:2009th,Benini:2009qs} (see \cite{Franco:2009sp} for related work as well). 

This theory admits an $\mathcal{N}=2$ supersymmetric AdS vacuum\footnote{See for instance \cite{Erbin:2014hsa} for further models of gauged $\mathcal{N}=2$ supergravity coupled to hypermultiplets with fully supersymmetric vacua.}. The field content of the theory is the gravity multiplet, two vector multiplets ($n_v=2$) and the universal hypermultiplet. We essentially follow the conventions of \cite{Gauntlett:2009zw,Cassani:2012pj}\footnote{In the original paper \cite{Gauntlett:2009zw} the vector kinetic terms have a factor $1/2$ instead of $1/4$ in front. However, their definition of $\mathcal{N}_{IJ}$ includes a factor $1/2$ with respect to ours, hence the total factor $1/4$ in our Lagrangian. These conventions differ with respect to those of \cite{Andrianopoli:1996cm} and \cite{Halmagyi:2013sla} by the following: $ A_{here} = \sqrt2 A_{there}$ and $k_{\Lambda}^u = \frac{1}{\sqrt2} k_{\Lambda}^u$, as already noticed (see footnote (10) of \cite{Cassani:2012pj}).}. The Lagrangian has the form
\begin{equation}
\label{N2action}
\begin{aligned}
   S =  \int \tfrac{1}{2}R\ast 1
      +{}& g_{i\bar{\jmath}}Dt^i\wedge\ast D\bar{t}^{\bar{\jmath}}
      + h_{uv}Dq^u\wedge\ast Dq^v \\
      & + \tfrac{1}{4}\im\mathcal{N}_{\Lambda \Sigma}F^{\Lambda}\wedge\ast F^{\Sigma}
      + \tfrac{1}{4}\re\mathcal{N}_{\Lambda \Sigma}F^{\Lambda}\wedge F^{\Sigma}
      - V \ ,
\end{aligned}
\end{equation}
where $t^i = \tau^i + i b^i, (i = 1,2)$ parameterize the two complex scalars in the vector multiplets and $q^u, (u=1, \ldots 4)$ those in the hypermultiplet. The vectors $F^\Lambda, (\Lambda = 0,1,2)$ come from the two vector multiplets and the gravity multiplet. We work in the symplectic frame where all gaugings are electric\footnote{This is the four-dimensional  theory obtained upon reduction, after dualization of the massive tensor multiplet in a massive vector multiplet (full details in \cite{Gauntlett:2009zw}).}, and the model is characterized by the corresponding holomorphic prepotential
\begin{equation}
  F(X) = -2i\sqrt{X^0 ( {X^1})^2 X^2} .
\end{equation}
The scalars in the vector multiplets parameterize the special K\"ahler manifold $\left(\frac{SU(1|1)}{U(1)} \right)^2$ with metric
\begin{align}
  g_{i \bar{\jmath}} &= \partial_i \partial_{\bar{\jmath}} \mathcal{K} (z, \bar{z}) \ , & \mathcal{K} &= - \log[i( \overline{X}^{\Lambda} F_{\Lambda} - X^{\Lambda} \overline{F}_{\Lambda})] \ .
\end{align}
where $F_{\Lambda} = \partial_{\Lambda} F$. From the covariantly holomorphic sections $(X^{\Lambda}, F_{\Lambda})$ we define moreover the sections
\begin{equation} \label{Llambda}
(L^{\Lambda}, M_{\Lambda}) = e^{\mathcal{K}/2} (X^{\Lambda},F_{\Lambda})\,.
\end{equation}
We choose sections such that $ X^{\Lambda} =\{X^0,X^1,X^2\}= \{1,t_1^2,t_2^2 \}$. The period matrix $\mathcal{N}_{\Lambda \Sigma}$ encodes the (scalar dependent) kinetic terms for the vector fields, and it is obtained via the special geometry relation
\begin{equation}
\mathcal{N}_{\Lambda \Sigma}= \overline{F}_{\Lambda \Sigma} + 2i \frac{\text{Im} F_{\Lambda \Delta} \text{Im} F_{\Sigma \Gamma} X^{\Delta} X^{\Gamma}  }{ \text{Im} F_{\Delta \Gamma} X^{\Delta} X^{\Gamma}}  \ ,
\end{equation}
where $F_{\Delta \Sigma} = \frac{\partial F}{\partial X^{\Delta} X^{\Sigma} }$. Its explicit form is reported in Appendix \ref{sec:conventions}. 

The universal hypermultiplet contains the 4 hyperscalars $q^u = (\phi,a,\xi ,\bar{\xi})$, which parameterize the quaternionic K\"ahler manifold $\frac{SU(2,1)}{S(U(2) \times U(1))}$ with metric $h_{uv}$ of the form
\begin{equation}
\label{univhyper}
   h_{uv} \td q^u \td q^v = \td \phi^2
       + \frac{e^{4 \phi}}{4} \left[
          \td a- \frac{i}{4}(\xi \td \bar{\xi}-\bar{\xi} \td \xi)\right]^2
       + \frac{e^{2\phi}}{4} \td \xi \td \bar{\xi} \; .
\end{equation}
This quaternionic K\"ahler manifold has constant negative curvature $R_q = -24 = -8 n_h (2 +n_h)$ \cite{Bagger:1983tt}, where $n_h$ is the number of hypermultiplets in the theory (in our case $n_h=1$). 

The covariant derivatives for the vector multiplets and the hyperscalars are given by
\begin{align}
D t^i &= \td t^i + k_{\Lambda}^i A^{\Lambda} \ , & D q^u &= \td q^u + k_{\Lambda}^u A^{\Lambda}\,,
\end{align}
where $k_{\Lambda}^i$ and $k_{\Lambda}^u$ are the Killing vectors corresponding to the gauging of the special K\"ahler and the quaternionic manifold respectively. The quaternionic Killing vectors $k_{\Lambda}^u$ can be derived from the Killing prepotentials $P_{\Lambda}^x$ which satisfy the relation $\Omega_{vw}^x k_{\Lambda}^w = - \nabla_{v} P_{\Lambda}^x$ \cite{Galicki:1986ja,Andrianopoli:1996cm,Andrianopoli:1996vr}, where $\Omega_{vw}^x = d \omega^x + \frac12 \epsilon^{xyz} \omega^y \wedge \omega^z $ is the curvature on the quaternionic manifold. In the model we consider, only a  U(1) isometry of the hypermultiplet manifold is gauged. Thus, the covariant derivatives for the vector multiplet scalars boil down to simple derivatives, as $k_{\Lambda}^i =0$. The hyperscalars are charged, however. The prepotentials and Killing vectors of the gauging are \cite{Gauntlett:2009zw,Cassani:2012pj}: 
\begin{align}
P_0 &= 6 P_{a} -4 P_{\xi}\,, & P_1 &= 4 P_{a}\,, & P_2 &= 2 P_a \,,
\end{align}
where
\begin{align}
P_a &= \left( \begin{array}{cc}
\frac{i e^{2 \phi}}{4} & 0 \\
0 & - \frac{i e^{2 \phi}}{4}
\end{array}
\right) \,, & P_{\xi} &= \left( \begin{array}{cc}
\frac{i}{2} (1- \xi \bar{\xi} e^{-2\phi}) & -i \xi e^{-\phi} \\
-i \bar{ \xi} e^{-\phi} & -\frac{i}{2} (1- \xi \bar{\xi} e^{-2\phi}) 
\end{array}
\right)
\end{align}
and $P_{\Lambda} = P_{\Lambda}^x \left( -\frac{i}{2} \sigma^x \right)$. Therefore,
\begin{align}\label{killingv}
k_0 & = -6 \partial_a +4i (\xi \partial_{\xi} - \bar{\xi} \partial_{\bar{\xi}}) \,, & k_1 & = -4 \partial_a \,, & k_2 &= -2 \partial_a \,.
\end{align}
Finally, the scalar potential of the theory, which couples scalars in the vector multiplets and hyperscalars, is given by
\begin{equation} \label{pot_unspec}
V (t, \bar{t}, q)= (g_{i \bar{\jmath}} k_{\Lambda}^i k_{\Sigma}^{\bar{\jmath}}+4h_{u v} k^u_{\Lambda} k_{\Sigma}^v) \bar{L}^{\Lambda} L^{\Sigma}+ (f_{i}^{\Lambda} f_{\bar{\jmath}}^{\Sigma} g^{i \bar{\jmath}} - 3 \bar{L}^{\Lambda} L^{\Sigma}) P_{\Sigma}^x P_{\Lambda}^x \,.
\end{equation} 
where $L^{\Lambda}$ are defined in \eqref{Llambda} and $f_i^{\Lambda} = (\partial_i +\tfrac12 \partial_i \mathcal{K} )L^{\Lambda}$.

Given this specific form of the gauging in the $M^{111}$ truncation, one of the vectors becomes massive via the Higgs mechanism. The spectrum then contains (see Table 7 of \cite{Cassani:2012pj})
\begin{itemize}
\item the gravity multiplet, containing the metric $g_{\mu \nu}$ and a massless vector,
\item a Betti vector multiplet, containing the massless vector and a complex scalar (two real fields) of mass $m^2 l^2 = -2$ (in our conventions, the Breitenlohner-Freedman bound is $m^2 l^2 \geq -9/4 $), each with $\Delta =(2,1)$,
\item a massive vector multiplet, containing a massive vector of mass $m^2 l^2=12$ (which corresponds holographically to a vector operator with weight $\Delta =5$), which has eaten its axion $a$ and five scalars of mass $m^2 l^2=(18,10,10,10,4)$ corresponding to $\Delta=(6,5,5,5,4)$.
\end{itemize}

Before proceeding further, let us remind the reader about the asymptotic fall-off of vectors and scalars in AdS$_4$ space-time. The scaling dimension of an operator dual to a massive $p$-form in AdS$_4$ space-time is given by the formula \cite{Aharony:1999ti}
\begin{equation}
\Delta_{\pm} =\frac32 \pm \frac12 \sqrt{(3-2p)^2+4m^2 l^2}\,.
\end{equation}
A vector field ($p=1$) dual to an operator of scaling dimension $\Delta$ behaves as ($r$ is the AdS radial coordinate, and the boundary is reached at $r \rightarrow \infty$)
\begin{equation} \label{falloff_vect}
r^{-2+\Delta_+} \qquad \text{and} \qquad r^{1-\Delta_+} \,.
\end{equation}
A scalar field ($p=0$) instead behaves as 
\begin{equation} \label{falloff_scalar}
r^{-3+\Delta_+}\qquad \text{and} \qquad r^{-\Delta_+} \,.
\end{equation}
We will come back to these asymptotic fall-offs later on when dealing with the explicit AdS$_4$ solutions.

\subsection{Consistent truncation}
\label{sec:consistentTruncation}

In finding black hole solutions we will make a simplifying assumption: we retain only \emph{one} hyperscalar. Indeed one can see that the complex hyperscalar $\xi$ can be consistently truncated away, and the field $a$ is the Stueckelberg field which can be consistently set to the value zero by a choice of gauge. Our truncated theory will then be characterized by the following matter content: two massless vector fields, a massive one, and five scalars of masses $m^2 l^2 = (18,10,4,-2,2)$ which correspond to dual operators of dimensions $\Delta = (6,5,4,(2,1),(2,1))$ where $(2,1)$ indicates the two normalizable modes for a scalar with mass $m^2 l^2 = -2$.

Given this truncation, the only nonvanishing components of the quaternionic Killing prepotentials are
\begin{equation} 
P_{\Lambda}^3 = (4-3 e^{2\phi}, -2 e^{2\phi}, -e^{2\phi} )\,,
\end{equation}
hence the Killing vectors appearing in the gaugings \eqref{killingv} are \begin{equation} \label{kv}
k_{\Lambda}^{a} = -(6, 4,2)\,.
\end{equation}  
In order to simplify our computation, we can assume a specific value for the Freund--Rubin parameter appearing in \cite{Halmagyi:2013sla}, $e_0 =6$, which leads to the fixed value of AdS radius  $l = \tfrac12 \left(\frac{e_0}{6} \right)^{3/4}=1/2$ - see for instance formula (3.16) of \cite{Halmagyi:2013sla}.

\bigskip

Putting all gauging data together, and redefining the hypermultiplet field $\phi = \log \sigma$, the action \eqref{N2action} can be rewritten in the form
 \begin{equation} \label{eq:action}
S = \int d^4x \sqrt{-g} \left(\frac12 R - V \right) + S_V +S_H\,,
\end{equation}
where the scalar potential is, using \eqref{pot_unspec},
\begin{equation}
V = \sigma^4 \left( \frac{(2 b_1 b_2 + b_1^2 + 3)^2}{\tau_1^2 \tau_2} + \frac{2(b_1 + b_2)^2}{\tau_2} + \frac{4 \tau_2 b_1^2}{\tau_1^2} + \frac{\tau_1^2}{\tau_2} + 2 \tau_2 \right) - 8 \sigma^2 \left(\frac{2}{\tau_1}+\frac{1}{\tau_2}\right) \ .
\end{equation}
It has an AdS minimum $V_\text{min} = -12$ for the following values of the scalar fields
\begin{equation} \label{vacua_scalar}
\tau_1=\tau_2=\sigma=1\,, \qquad b_1=b_2=0\,.
\end{equation}
The action for the hypermultiplet sector is
\begin{equation}
  S_H = -\frac{1}{2}\int{\td^4 x \sqrt{-g} \left[2 \big(\nabla \log \sigma \big)^2 + \frac12 \sigma^4\big(\nabla a - (6 A_0 +4 A_1+2 A_2)\big)^2 \right]} \ ,
\end{equation}
where we can see that the scalar field $a$ acts as a Stueckelberg field responsible for the Higgsing of the linear combination $6 A_0 +4 A_1+2 A_2$. Finally, the vector multiplet Lagrangian reads
\begin{align}
  S_V &\eq \frac14 \int{\td^4 x \sqrt{-g} \left[ -2\left( \nabla(\log \tau_1) \right)^2 - \left( \nabla(\log \tau_2) \right)^2 - \frac{2(\nabla b_1)^2}{\tau_1^2}  -\frac{(\nabla b_2)^2}{\tau_2^2}  \right]} \nonumber \\
  &\pheq + \frac14 \int{\left( \im \cN_{\Lambda \Sigma} F^\Lambda \wedge *F^\Sigma + \re \cN_{\Lambda \Sigma} F^\Lambda \wedge F^\Sigma \right)} \ ,
\end{align}
with $\cN$ given in \eqref{eq:cN}. The supergravity vector fields $A^{\Lambda}$ can be expressed as well as linear combination of the massless eigenstates $\mathcal{A}_1$,  $\mathcal{A}_2$ and the massive one $\mathcal{B}$, in this way
\begin{eqnarray} \label{eigen_vec}
A^0 & =& \frac{1}{2}\mathcal{A}_1 +\frac{\sqrt3}{2} \mathcal{B}\,, \nonumber \\
A^1 & =& -\frac{1}{2} \mathcal{A}_1 + \frac{\sqrt3}{6} \mathcal{B}- \frac{1}{\sqrt6} \mathcal{A}_2 \,, \nonumber \\
A^2 & =& -\frac{1}{2} \mathcal{A}_1 + \frac{\sqrt3}{6} \mathcal{B}+ \frac{2}{\sqrt6} \mathcal{A}_2\,.
\end{eqnarray}
We checked that this action reduces to that of \cite{Gauntlett:2009zw} if we identify $t_1 = t_2$ and $A^1=A^2$. These identifications correspond to switching off the Betti vector multiplet, which contains in particular the massless vector $\mathcal{A}_2$. The universal $SE^7$ reduction of \cite{Gauntlett:2009zw} coincides with the truncation  on  $S^7 = SU(4)/SU(3)$ that retains the $SU(4)$ left-invariant modes.

%% file: Solutions.tex
\subsection{Static black hole ansatz}\label{static_BH_ansatz}

We focus on the search for static and spherically symmetric solutions of the form\footnote{We look for configuration of spherical horizon topology but we expect that solution with flat or hyperbolic event horizons exist as well, as BPS solutions of this kind were found in \cite{Halmagyi:2013sla,Donos:2012sy}.}
\begin{equation} \label{metricR}
ds^2 = -e^{-\beta(r)} h(r) \, dt^2 + \frac{dr^2}{h(r)} +r^2 \, d\Omega^2 \,,
\end{equation}
which allows for asymptotically locally AdS space-times. The five scalar fields have only radial dependence:
\be \label{scalarR}
\tau_1 = \tau_1(r)\,, \qquad \tau_2 = \tau_2(r) \,, \qquad b_1=b_1(r)\,, \qquad b_2=b_2(r)\,, \qquad \sigma=\sigma(r)\,.
\ee
For the vectors, we choose an ansatz that can describe the fields around a static black hole with both electric and magnetic charge,
\begin{align}\label{ans_vec}
\mathcal{A}_{1,t}  &= \xi_1(r) \,, & \mathcal{A}_{2,t} & = \xi_2 (r)\,, & \mathcal{B}_t & = \zeta (r)\,, \nonumber \\
 \mathcal{A}_{1,\varphi} &= P^{1} \cos \theta\,, & \mathcal{A}_{2,\varphi} &= P^{2} \cos \theta\,, & \mathcal{B}_{\varphi} &= P^m \cos \theta\,.
\end{align}
More precisely, the charges are the integral of the flux of the field strength $F_{\mu \nu}$ and its dual $G_{\mu\nu}$ through the sphere at spatial infinity:
\begin{equation}\label{chargesflux}
Q_{i} = \frac{1}{4 \pi} \int_{S^2_\infty} G_{\mathcal{A}_i} \,, \qquad  P^{i} =  \frac{1}{4 \pi} \int_{S^2_{\infty}}  F_{\mathcal{A}_i}\,,
\end{equation}
with the dual defined as
\begin{equation} \label{Fdual}
G_{\mu \nu, \Lambda} = \frac14 \sqrt{-g} {\epsilon_{\mu \nu}}^{\rho \sigma} \frac{\partial \mathcal{L}}{\partial F^{\rho \sigma, \Lambda}} \,.
\end{equation}
The equations of motion derived from \eqref{N2action} with the above ansatz are given in Appendix \ref{sec:EoM}. In total, there are 14 degrees of freedom: the equations of motion for the metric components $\beta$ and $h$ are first order, yielding one dynamic component each. The scalars $\tau_1$, $\tau_2$, $b_1$, $b_2$ and $\sigma$ on the other hand, have second order equations of motion. Just like the massive vector mode $\zeta$. Due to charge conservation and gauge invariance, there are no dynamical components that correspond to the massless vectors $\xi_1$ and $\xi_2$. 

In the duality frame we consider, the hypermultiplets and the gravitini are electrically charged. Therefore, the following Dirac quantization conditions need to hold:
\begin{equation} \label{dirac}
P^{\Lambda} k_{\Lambda}^u(\bar{q}) \in \mathbb{Z}\,,
\qquad
P^{\Lambda} P_{\Lambda}^3(\bar{q})  \in \mathbb{Z}\,,
\end{equation}
where $P_{\Lambda}^3 (\bar{q}) = \{1, -2 , -1 \}$ and $k_{\Lambda}^u(\bar{q})=-\{6, 4 , 2 \}$ are respectively the Quaternionic  Killing prepotentials and Killing vectors computed on the vacuum solution \eqref{vacua_scalar}. The first Dirac quantization condition in \eqref{dirac} is automatically satisfied on shell for the particular assumptions on the ansatz we made, while the second condition in \eqref{dirac}, taking into account \eqref{eigen_vec}, reads
\begin{equation}
2 P^1 \in \mathbb{Z} \,.
\end{equation}
Furthermore, the Maxwell equation \eqref{maxwell_equation} imposes to the condition
\begin{equation}
P^{\Lambda} k_{\Lambda}^u =0
\end{equation}
along the entire flow. In our case this means that the massive vector in \eqref{ans_vec} field has zero magnetic component:
\be
P^m=0\,.
\ee
Releasing the condition of spherical symmetry would allow for a nontrivial magnetic component. In particular, this would result in vortex lines of the  Nielsen-Olsen \cite{Nielsen:1973cs} type\footnote{The strings stretched between probes mentioned in the introduction would manifest themselves as vortex-type solutions in this kind of truncation. This would be interesting to study, but it goes beyond the scope of the present work. We hope to come back to this point in the future.}. 

Similarly to \cite{Gauntlett:2009dn,Bobev:2011rv}, the equation of motion and the background fields have the following scaling symmetry, 
\be
t \rightarrow \gamma t  \ , \qquad \beta \rightarrow \beta + 2 \log \gamma \ , \qquad \xi_1 \rightarrow \frac{\xi_1}{\gamma} \ , \qquad  \xi_2 \rightarrow \frac{\xi_2}{\gamma} \ , \qquad \zeta \rightarrow \frac{\zeta}{\gamma}  \ ,
 \ee
which can be used to choose without loss of generality the asymptotic value of the metric function $\beta$ at infinity. Indeed  in what follows we will choose 
\be
\lim_{r \rightarrow \infty} \beta  =0\,.
\ee

The black hole solutions are most conveniently represented by the coordinate $u$, which is related to the radial Schwarzschild coordinate as
\begin{equation}
u= \log \left(\frac{r}{r_H} \right) \,,
\end{equation}
where $r_H$ is the location of the event horizon. 
The horizon is retrieved by the $u=0$ limit, while asymptotically $u \rightarrow \infty$ the solution approaches AdS$_4$ space-time, with radius $l_{AdS}=2$, which is kept fixed in our computations\footnote{It is nevertheless straightforward to reinstate the gauge coupling constant in the action \eqref{N2action}, allowing for a different value of the cosmological constant and $AdS$ radius. The authors of \cite{Gauntlett:2009dn,Bobev:2011rv} moreover find another scaling symmetry which allows to pick $r_H$ =1 without loss of generality. This is due to the fact that they deal with planar horizons - in case of spherical horizons such additional scaling symmetry (3.23) of \cite{Bobev:2011rv} is absent, as one can see comparing our equation \eqref{einsteintt} with and (2.17) of \cite{Bobev:2011rv}.}. In these new coordinate $u$, the metric reads:
\begin{equation}
ds^2=- e^{2u-\beta(u)} \,\, r_H^2 \,\, H(u) \, dt^2 +\frac{du^2}{H(u)}+ e^{2u}\,\, r_H^2(d\theta^2 + \sin^2 \theta d \phi^2)\,,
\end{equation}
where we defined 
\be
h(u) = r_H^2 \, e^{2u} \,H(u)\,.
\ee

As an elementary consistency check, the dyonic Reissner-Nordstr\"om solution is obtained by setting all the scalar fields at their vacuum value \eqref{vacua_scalar} throughout the entire flow. The solution is then characterized by the following warp factors:
\begin{align}
\beta &= 0 \ , &  H(u) &= 4+ \frac{e^{-2 u}}{r_H^2}- \frac{(16 r_H^4 + 4r_H^2 + (P^1)^2 + Q_1^2) \, e^{-3u} }{4r_H^4} +\frac{((P^1)^2+Q_1^2) \, e^{-4 u}}{4 r_H^4}\,, \label{hrn}
\end{align}
with the additional conditions
\be
P^2 =0\,, \qquad Q_2=0\,.
\ee
coming from the scalar equations of motion. The Reissner-Nordstr\"om solution in \eqref{hrn} is parameterized by the electromagnetic charges $Q_1$ and $P^1$ and the radius of the event horizon $r_H$ which can be equivalently traded for the mass $M$ of the black hole.

\subsection{Strategy for numeric simulations}

We will use numerical tools to solve the equations of motion subject to the relevant boundary conditions. This allows us to find smooth configurations which in the UV approach AdS space-time and in the IR form the black hole horizon. In order to preserve the AdS$_4$ asymptotics we need to set the diverging modes of the heavy scalars and of the massive vector field to zero (see eq. \eqref{falloff_vect}-\eqref{falloff_scalar}). The requirement of regularity on the black hole horizon will relate the derivative of the scalar fields to their values at the horizon.

To find solutions interpolating between AdS$_4$ and the black hole horizon, we will use a shooting method. In a first step, we provide boundary conditions in the IR, i.e. on the black hole horizon at $u = 0$, and integrate the equations of motion towards the boundary of AdS. Secondly and independently, we choose boundary conditions in the UV (at a value $u \gg 1$, so $r \gg l$) and integrate the equations into the IR. At some intermediate point in the bulk (for example $u = 1$), we obtain two values for each of the fields, depending nonlinearly on both sets of boundary conditions we chose. We then employ an optimization algorithm to minimize the difference and finally obtain the matching, by tweaking the boundary conditions on both the black hole horizon and the asymptotic boundary of AdS. 

As mentioned before, there are 14 dynamic degrees of freedom. As we will see below, we can tune 16 boundary conditions for the fields\footnote{It turns out that for each field except for the scalars with $m^2 l^2 = -2$, the conditions of a smooth black hole horizon and asymptotically AdS fix as many boundary conditions as there are degrees of freedom. The light scalars have two normalizable modes for $r \to \infty$, both of which are compatible with the asymptotic AdS behavior.}, as well as the value of $r_H$ and four black hole charges (two electric and two magnetic). We therefore expect to find a 7-parameter family of solutions\footnote{\label{foot:uniqueness}Due to the nonlinear nature of this system, this naive expectation is possibly incorrect. In principle, there might be no solutions at all, or there could be multiple 7-parameter families of solutions, up to a countable infinite number of them.}.

\subsection{Asymptotic behavior of the fields}

The solution near the AdS boundary is characterized by the fall-off of the fields. They are most conveniently expressed in terms of the coordinate $z = 1/r = e^{-u}/r_H$. The most general boundary conditions can be derived by considering the equations of motion order by order in $z$, starting from the leading terms which are fixed by the requirement of asymptotic AdS. To obtain the most general solution, one should take into account terms of the form $z^n$ as well as $z^n \log(z)^m$ (see for example \cite{Halmagyi:2013sla}). The equations of motion will require most (but not all) of the logarithmic terms to vanish. In total, we find 9 coefficients $(h_{(3)}, \tau_{(1)}, b_{(1)}, \tau_{(2)}, b_{(2)}, \sigma_{(4)}, b_{(5)}, \sigma_{(6)}, \zeta_{(4)})$, which encode the asymptotic behavior as follows.

\begin{subequations} \label{exp_infinity}
  The components of the metric have the following fall-off:
  \begin{align}
    H &= 4 + \left(1 + 6 \tau_{(1)}^2 + 6 b_{(1)}^2 \right) z^2 + h_{(3)} z^3 + \order{z^4} \nonumber \\
    \beta & = \frac32 \left( \tau_{(1)}^2 + b_{(1)}^2 \right) z^2 + \frac45 \left( \tau_{(1)}^3 - \tau_{(1)} b_{(1)}^2 + 5\tau_{(1)} \tau_{(2)} + 5 b_{(1)} b_{(2)} \right) z^3  + \order{z^4} \ , \label{eq:metricFallOff}
  \end{align}
  where $\tau_{(1)}$ and $b_{(1)}$ are the leading fall-off coefficients of the lightest scalar fields (see below). If they vanish, we recover the familiar AdS-Reissner-Nordstr\"om with $M = -h_{(3)} / 2$ as in \eqref{hrn}. As mentioned before, we choose the time coordinate such that $\beta|_{z=0} = 0$.
  
  The behavior of the scalar fields can be expressed as a power series in $z \sim 0$ as well (for the sake of clarity, we omit terms that are at least quadratic in the coefficients)
  \begin{align} \label{eq:scalarExpansion}
    \tau_1 & = 1 + \tau_{(1)} z + \tau_{(2)} z^2 + \ldots + \left( \frac43 \sigma_{(4)} - \frac1{12} \tau_{(2)} + \ldots \right) z^4 + \ldots \nonumber \\
    &\pheq - \left( \sigma_{(6)} + \frac12 \sigma_{(4)} + \frac1{80} \tau_{(2)} + \ldots \right) z^6 + \order{z^7}\,, \nonumber \\
    \tau_{2} &= 1 - 2 \tau_{(1)} z - \left( 2\tau_{(2)} + \ldots \right) z^2 + \ldots + \left( \frac43 \sigma_{(4)} + \frac16 \tau_{(2)} + \ldots \right) z^4 \nonumber \displaybreak[0] \\
    &\pheq + \ldots - \left( \sigma_{(6)} + \frac12 \sigma_{(4)} + \frac1{40} \tau_{(2)} + \ldots \right) z^6 + \order{z^7} \,, \nonumber \displaybreak[0] \\
    b_1 & = b_{(1)} z + b_{(2)} z^2 + \ldots - \left( \frac1{12} b_{(2)} + \ldots \right) z^4 + (b_{(5)} + \ldots) z^5 + \order{z^6} \,,\nonumber \displaybreak[0] \\
    b_2 & = -2 b_{(1)} z + (3 b_{(1)} \tau_{(1)} - 2 b_{(2)}) z^2 + \ldots + \left( \frac16 b_{(2)} + \ldots \right) + (b_{(5)} + \ldots) z^5 + \order{z^6} \,, \nonumber \\
    \sigma &= 1 + \ldots + \left( \sigma_{(4)} + \ldots \right) z^4 + \ldots + \left( \sigma_{(6)} + \ldots \right) z^6 + \order{z^7}\,.
  \end{align}
  To zeroth order in $z$, the scalars are in the AdS extremum of the potential \eqref{vacua_scalar}. As anticipated in Section \ref{sec:consistentTruncation}, the excitations around this minimum are characterized by the eigenvalues $m^2 l^2 = (18, 10, 4, -2, -2)$ of the mass matrix. Therefore, there are two independent components of the fields with fall-off $z$ (corresponding to a $\Delta = 1$ source or operator in the CFT, depending on the quantization scheme), parameterized by $\tau_{(1)}$ and $b_{(1)}$; there are two modes falling off like $z^2$, proportional to $\tau_{(2)}$ and $b_{(2)}$; and there are single modes proportional to $z^4$, $z^5$ and $z^6$, parameterized by $\sigma_{(4)}$, $b_{(5)}$ and $\sigma_{(6)}$, respectively. Furthermore, interactions give rise to terms quadratic in these coefficients, which are included in the ``$\ldots$''.
  
  Finally, the massive vector field $\zeta$ has the following fall-off
  \begin{align}
  \label{eq:vectorExpansion}
    \zeta &= \frac{\sqrt{2}}{10} \left( Q_2 \tau_{(1)} - b_{(1)} P^2 \right) z^2 + \frac{\sqrt{2}}{3} \left( Q_2 \tau_{(2)} - b_{(2)} P^2 + \ldots \right) z^3 \nonumber \\
    &\pheq - \frac{3 \sqrt{2}}{70} \left( Q_2 \tau_{(1)} - b_{(1)} P^2 + \ldots \right) z^4 \log(z) + \zeta_{(4)} z^4 + \order{z^5}\,.
  \end{align}
  The presence of the massive vector on the gravity side signals a broken global flavor symmetry in the dual field theory. The parameter $\zeta_{(4)}$ is related to the expectation value of a dual operator with dimension $\Delta=5$.
\end{subequations}

\subsection{Fields at the horizon}

The boundary conditions on the black hole horizon, which in our conventions is located at $u=0$, must ensure the existence of a smooth horizon. The timelike component of the metric $g_{tt} \propto H$ must vanish while none of the scalar fields must diverge. Furthermore, consistency of the equation of motion requires the massive vector field $\zeta(u)$ to vanish (leaving only its derivative as a free parameter) and determine the derivatives of the scalar fields in terms of their values at the horizon. All together, the fields near $u \approx 0$ are characterized by 7 parameters $(\beta^{(h)}, \sigma^{(h)}, \tau_1^{(h)}, \tau_2^{(h)}, b_1^{(h)}, b_2^{(h)}$, $\zeta'^{(h)})$,
\begin{align} 
H &= u \left( 12 + \frac1{r_H^2} + \frac{1}{12 r_H^4} \left[ -3 \left( Q_1^2 + Q_2^2 + (P^1)^2 + (P^2)^2 \right) - \left( Q_2^2 - (P^2)^2 \right) (4 \tau_1^{(h)} - \tau_2^{(h)}) \right. \right. \nonumber \\
&\pheq \left. \left. + 2 Q_2 P^2 (4b_1^{(h)} - b_2^{(h)}) + 2\sqrt{6} (Q_1 P^2 + Q_2 P^1) (b_1^{(h)} - b_2^{(h)}) \right. \right. \nonumber \\
&\pheq \left. \left. + 2 \sqrt{6} (Q_1 Q_2 - P^1 P^2) (\tau_1^{(h)} - \tau_2^{(h)}) \right] + \ldots \right) + \order{u^2} \ , \nonumber
\end{align}
\vspace{-20pt}
\begin{align} \label{exp_horizon}
\beta &= \beta^{(h)} + \order{u}  \ , & \tau_1 &= \tau_1^{(h)} + \order{u}  \ , & \tau_{2} &= \tau_2^{(h)} + \order{u}  \ , \\
b_1 &= b_1^{(h)} + \order{u}   \ , & b_2 &= b_2^{(h)} + \order{u} \ , & \sigma &= \sigma^{(h)} + \order{u}  \ , & \zeta &= \zeta'^{(h)} u + \order{u^2}  \ . \nonumber
\end{align}

\subsection{Solutions: examples}
\label{sec:examples}
At this point, there are 21 free parameters: 
\begin{itemize}
  \item 9 boundary conditions on the asymptotic boundary of AdS,
  \item 7 boundary conditions on the black hole horizon, 
  \item the radius of the event horizon $r_H$,
  \item the electromagnetic charges of the black hole $(Q_1,Q_2,P^1,P^2)$ which represent conserved quantities of the two massless vector fields.
\end{itemize}
To obtain a consistent AdS black hole solution, however, one cannot choose all of these parameters arbitrarily. There are 14 constraints from the requirement that the IR solution to the equations of motion (integrated from the black hole horizon outward) evolve smoothly into the UV solution (integrated from the boundary of AdS inward). Indeed, the equations of motion are a system of 14 coupled first order ODEs. Thus, 14 integration constants must be fixed to ensure a smooth solution. We collectively denote them by $q_{integr}$.

The system is then still underdetermined: we have $21 - 14 = 7$ tunable parameters which are not fixed by the equations of motion, which by themselves specify each black hole solution taken into consideration. These are the four electromagnetic charges $(Q_1,Q_2,P^1,P^2)$, the leading modes of the light scalar fields $\tau_{(1)}$ and $b_{(1)}$, and the radius of the event horizon $r_H$. We denote these parameters by $q_{input}$.

With this in mind, one can find solutions numerically. We developed a Mathematica code that, given a set of external tunable parameters $q_{input}$, allows us to find black hole solutions by finding appropriate $q_{integr}$. The results are fully backreacted configurations representing thermal black hole solutions with nontrivial radial profile for the matter present in the theory.

\begin{figure}[htbp]
  \centering
  \includegraphics[scale=.7]{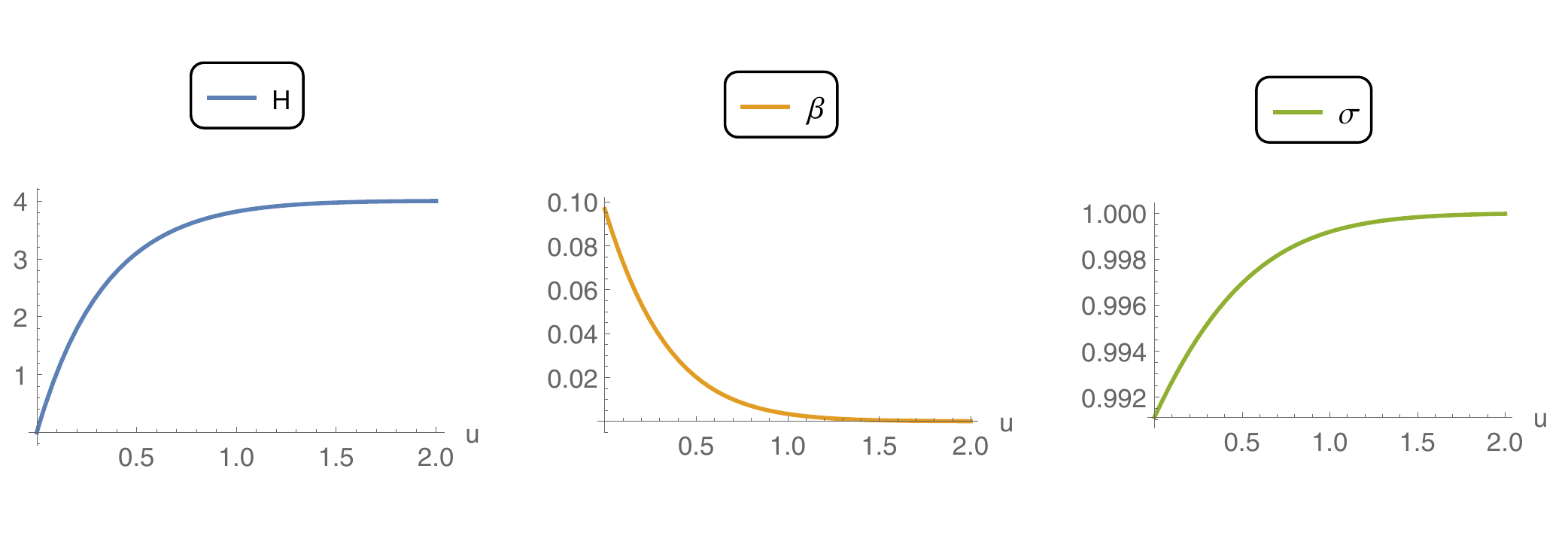}
  \includegraphics[scale=.7]{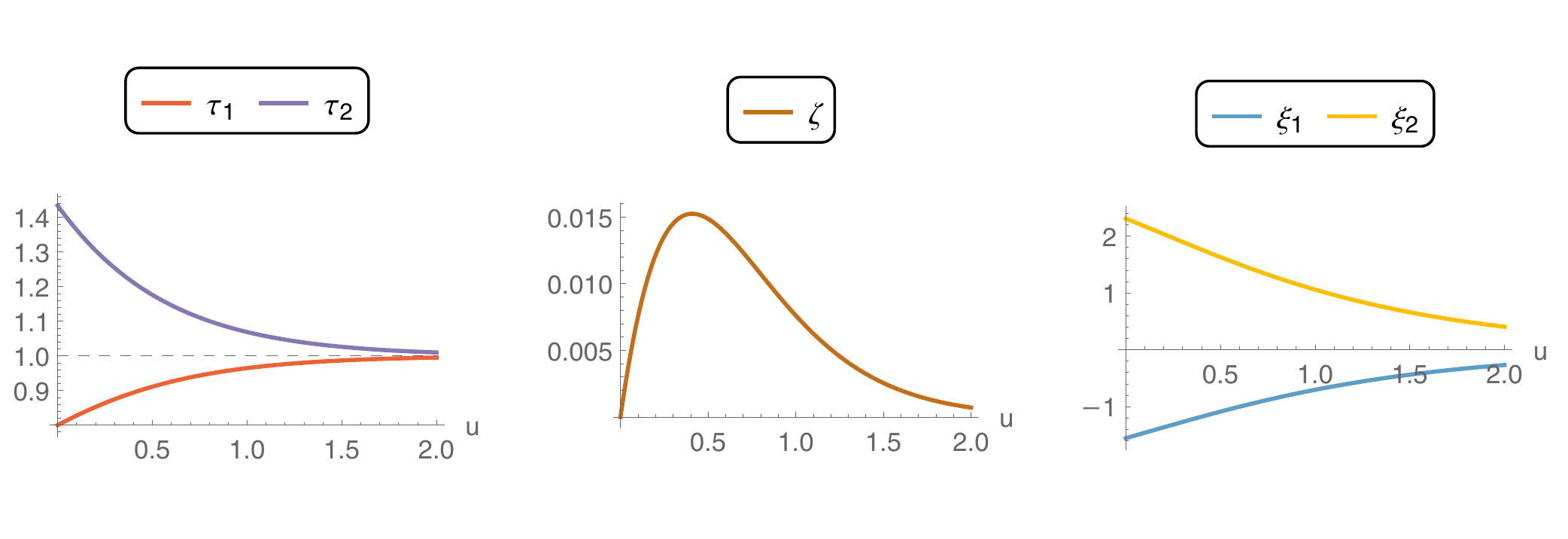}
  \caption{Purely electric black hole solution with $r_H = 1$, $Q_1 = 2$, $Q_2 = -3$, $P^1 = P^2 = 0 = \tau_{(1)} = b_{(1)}$. The integration constants obtained with the numerical shooting technique are (we report them here up to 3 digits) $\beta^{(h)}=0.097, \,\, \sigma^{(h)}=0.991, \,\,  \tau_1^{(h)} = 0.800, \,\, \tau_2^{(h)} = 1.434, \,\,b_1^{(h)}= b_2^{(h)}= 0, \,\,{\zeta'}^{(h)}=0.092, \,\, h_{(3)}= -7.816,\,\,\tau_{(1)}=0,\,\, \tau_{(2)} =-0.270,\,\, b_{(1)}=b_{(2)} =b_{(5)}=0,\,\, \sigma_{4} =-0.036, \,\, \sigma_{(6)}= -0.082,\,\, \zeta_{(4)}=-0.728$. The IR solution was integrated from $u = 10^{-12}$ to $u = 1$, and the UV solution was integrated from $u = 10 \to u = 1$. We used 30 digits of numerical precision. The IR and UV solutions at $u = 1$ differ by $\sum_i (\Delta \varphi_i)^2 = 1.22 \cdot 10^{-23}$, where $\varphi = (H, \beta, \tau_1, \tau'_1, \tau_2, \tau'_2, b_1, b'_1, b_2, b'_2, \sigma, \sigma', \zeta, \zeta')$.}
  \label{figure_electric}
\end{figure}

\begin{figure}[htbp]
  \centering
  \includegraphics[width=\textwidth]{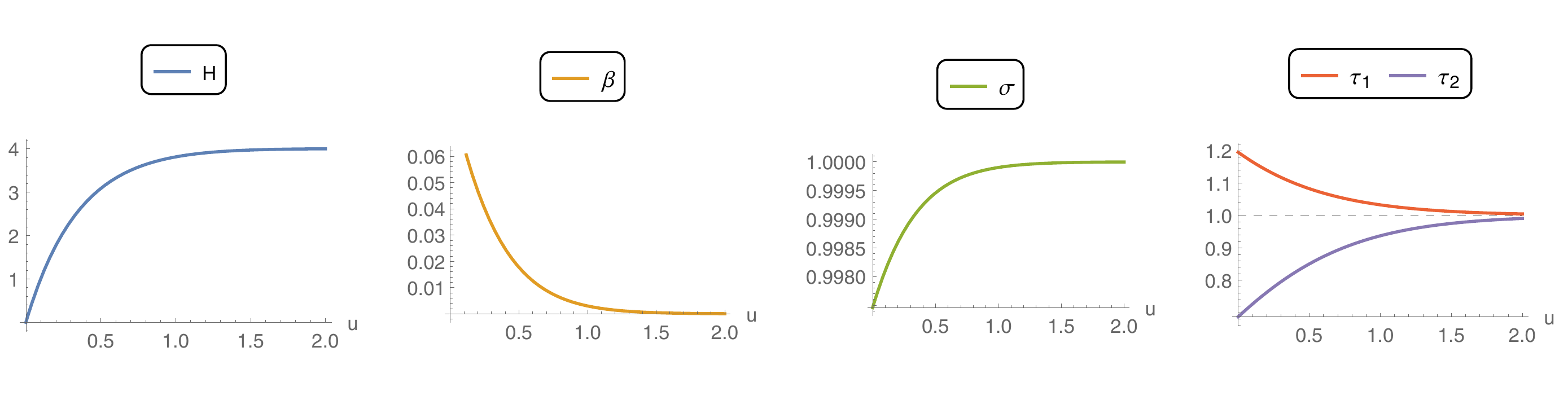}
  \caption{Purely magnetic black hole solution with $r_H = 1$, $P^1 = 2$, $P^2 = -3$, $Q^1 = Q^2 = 0 = \tau_{(1)} = b_{(1)}$. The massive vector $\zeta$, as well as $\xi_1$ and $\xi_2$ vanish identically in this case. The integration constants are $\beta^{(h)}=0.085, \,\, \sigma^{(h)}=0.997, \,\,  \tau_1^{(h)} = 1.195, \,\, \tau_2^{(h)} = 0.687, \,\,b_1^{(h)}= b_2^{(h)}= 0, \,\,{\zeta'}^{(h)}=0, \,\, h_{(3)}= -7.850,\,\, \tau_{(1)}=0,\,\,\tau_{(2)} =0.257,\,\, b_{(1)}= b_{(2)} =b_{(5)}=0,\,\, \sigma_{4} =-0.004, \,\, \sigma_{(6)}= 0.0229,\,\, \zeta_{(4)}=0$. We used 30 digits of numerical precision. The IR and UV solutions at $u = 1$ differ by $\sum_i (\Delta \varphi_i)^2 = 1.16 \cdot 10^{-23}$.}
  \label{figure_magnetic}
\end{figure}

We find electric, magnetic and dyonic solutions. The behavior of the fields as a function of the radial coordinate $u$ is displayed in Figure \ref{figure_electric} for the purely electric configuration, and in Figure \ref{figure_magnetic} for the purely magnetic one. In the latter case the massive vector is zero (see discussion in Section 3.1). 

The vector condensate surrounding the black hole and is moreover visualized in the 2d radial the plot in Figure \ref{Fig2d}. Circles of radius $r$ in the plot truthfully correspond to spheres with surface area $4\pi r^2$ in the AdS black hole geometry. However, radial distances in the plot are related to radial distances in the AdS black hole geometry by $\td r_\text{plot} = \td r_\text{BH} / \sqrt{H(r_\text{BH})}$, where $H$ is given in Figure \ref{figure_electric}. The massive vector field $\zeta$ vanishes at the event horizon, and it is peaked at a finite radial value outside the black hole horizon. The field $\zeta$ by itself is  a massive object surrounding the black hole: the configuration can therefore be seen as an example of ``composite'' back-reacted configuration in AdS space-time. The interactions dictated by the nontrivial couplings of the supergravity Lagrangian allow this massive object to gravitate outside the black hole horizon without falling in. 

It would be interesting to understand more deeply why the massive vector halo is stable outside the horizon. For example, one might attempt to analyze the stability of a ``probe'' massive vector particle in this background, in analogy with the probe black hole calculation of \cite{Anninos:2013mfa}. However, the point particle approximation can be expected to break down since the de Broglie wavelength of such a particle is of the order of the AdS length scale. Furthermore, there is kinetic mixing between the vectors in the supergravity Lagrangian, which is expected to affect the effective particle interactions. One would need to overcome these obstacles to obtain the correct form of the effective potential for the probe, and determine its stability. In relation to this, it would be interesting to investigate if configurations other than a black hole can support such massive vector field halo as well.

Before concluding, let us stress one difference between our solutions and those treated for instance in  \cite{Gubser:2008px,Hartnoll:2008vx,Hartnoll:2008kx,Denef:2009tp}. In our case the configuration has a massive vector in the Kaluza-Klein spectrum. Therefore the related symmetry is broken already in the vacuum of the theory, and it is never restored. However, for the solutions in \cite{Gubser:2008px,Hartnoll:2008vx,Hartnoll:2008kx,Denef:2009tp} describing holographic superconductors, the vector field in the vacuum of the theory has zero mass, as one can see from the asymptotic expansion of the fields. The breaking of the $U(1)$ symmetry happens in the latter case only in the proximity of the horizon, while the symmetry is restored at the boundary. One can actually see that the linearized theory considered for instance in \cite{Denef:2009tp} retains the $\mathcal{A}_1$ gauge field and the scalar mode $\xi$ that we instead truncated away. 

The configurations we find are also different from those in \cite{Horowitz:2014gva}, where black hole solutions hovering outside a black brane horizon were found. In this latter case the tendency of the object to fall towards the horizon is balanced by the electrostatic force towards the boundary due to a charged defect in the 3d dual CFT.

In addition, one of the massless vectors in our theory, the Betti vector, comes from internally wrapped branes, and it is dual to a baryonic symmetry in the dual field theory \cite{Benini:2009qs}. None of the $U(1)$ gauge fields in the others models we mentioned are dual to baryonic symmetries. We will revisit these points later when we deal with the phase transitions in the canonical ensemble.

%% file: FirstLaw.tex
We now proceed with the computation of the thermodynamic quantities of the black hole solutions we have found. The first law of black hole thermodynamics is related to the variation of the on-shell action, which by virtue of the equations of motion is a boundary term. In order to compute the Euclidean on-shell action, we first analytically continue the expressions \eqref{eq:action} and \eqref{metricR},
\be
I = -iS \,, \qquad t = -i \tau\,.
\ee
The Hawking temperature of the black hole (in units where the Boltzmann constant is 1) is
\be \label{temperature_expl}
T =  \frac{[h' e^{-\beta/2}]_{r=r_H}}{4\pi}\,,
\ee
and it is computed  by demanding regularity of the Euclidean geometry obtained from \eqref{metricR} at $r= r_H$. 
The Bekenstein-Hawking entropy is given by (with $G_N = 1$)
\be \label{BHentropy}
S = \pi r_H^2 
\ee
Furthermore, the electromagnetic charges were defined in \eqref{chargesflux} and the corresponding electrostatic and magnetostatic potentials are (in a gauge for which the vector potentials vanish on the boundary of AdS)
\begin{align}
  \phi^{\mathcal{A}_i} &\equiv - \int_{r_h}^{\infty} {F_{\mathcal{A}_i, tr} \, \td r} = \mathcal{A}_{i,t} (r_h) = \xi_i(r_h)  \ , & \chi_{\mathcal{A}_i} &\equiv - \int_{r_h}^{\infty} {G_{ \mathcal{A}_i, tr} \, \td r}  \ .
\end{align}
We will show in the next section that the mass of the AdS black hole solutions receives contribution from the scalar fields
\begin{equation}
M = -\frac{h_{(3)}}{2}+ 2(e_{a,2} \,e_{a,1}+ e_{b,2} e_{b,1})\,. \label{mass_eig}
\end{equation}
The quantities $e_{a,i}$, $e_{b,i}$ are the asymptotic falloff of the  $m^2 l^2 = -2$ eigenstates of the mass matrix
\begin{align}\label{as_fall2}
\phi_{-2,a} &= \frac{e_{a,1}}{r}+\frac{e_{a,2}}{r^2} + O(r^{-3})  \ , &  \phi_{-2,b} &= \frac{e_{b,1}}{r}+\frac{e_{b,2}}{r^2} +O(r^{-3}) \,.
\end{align}

The first law of thermodynamics relates these quantities along a family of black hole solutions, in this way:
\begin{align}
  \td M &= T \td S + \phi^{\mathcal{A}_i} \td Q_i \ . \label{eq:firstLaw}
\end{align}
This relation can be checked analytically for AdS Reissner-Nordstr\"om \eqref{hrn}. For our numerical solutions, it provides a nontrivial consistency check. Indeed, the thermodynamic quantities can be computed from the behavior of the fields either close to the black hole horizon or near the boundary of AdS. The relation \eqref{eq:firstLaw} indicates that they are not unrelated: they are correlated by the existence of a solution to the equations of motion that interpolates between these distant regions and is regular everywhere.

\subsection{Renormalized on-shell action}
Our Lagrangian contains a Higgsed vector field and scalar fields which are dual to irrelevant operators. Therefore, we must take care to identify the correct counterterms and obtain a finite result for the on-shell action. Holographic renormalization in presence of massive vector fields was worked out in \cite{Korovin2014}, where the necessary counterterms to renormalize the Proca-AdS action were obtained via the Hamilton-Jacobi formalism. Moreover, vector fields acquiring mass via spontaneous symmetry breaking were  considered in \cite{Bianchi:2001kw,Martelli:2002sp}.

As explained in the previous sections, the diverging modes for all but the lightest scalar fields must be required to vanish in order not to spoil the AdS asymptotics \footnote{One could however turn on the sources for these irrelevant operators perturbatively, as done for example in \cite{vanRees:2011fr}. We thank A. Bzowski, Y. Korovin and I. Papadimitriou for discussions about this point.}. In the Hamilton-Jacobi procedure for holographic renormalization \cite{deBoer:1999tgo} 
the vanishing of the diverging modes can be formulated as a set of second-class constraints, ensuring consistency (see for instance \cite{Chemissany:2014xsa,An:2016fzu}).
This means that, when deriving the equations of motion using the variational principle, the coefficients of the non-normalizable modes will be fixed to zero.

Provided these constraints are satisfied, the counterterms that renormalize the action are the Gibbons-Hawking term $I_{GH}$, the canonical counterterms $I_{ct}$ and the counterterm $I_{ct,A}$ due to the presence of the massive vector field as in \cite{Korovin2014}:
\begin{equation} \label{ren_os_action}
I_{ren} = I + I_{GH} + I_{ct}+ I_{ct,A}\,.
\end{equation}
The term $I_{GH}$ is of the form
\begin{align}\label{count_GH}
I_{GH} &=\frac12 \int_{\partial M}{\sqrt{g_3} \, \Theta} \ , \qquad \qquad \Theta_{\mu\nu} = -(\nabla_{\mu} n_{\nu} + \nabla_\nu n_\mu) \,.
\end{align}
where $g_{3,ab}$ is the induced metric on $\partial M$, $\Theta$ is the trace of the extrinsic curvature, and $n_{\mu}$ is the unit vector normal to the boundary. The term $I_{ct}$ contains the counterterms necessary to cancel the divergences \cite{Emparan:1999pm}
\begin{equation} \label{canon}
I_{ct}=\int_{\partial M} d^3x \sqrt{g_3} \left[\frac{l}{2} \mathcal{R} - \frac{l^3}{2} \left( \mathcal{R}_{bc} \mathcal{R}^{bc} -\frac{3 \mathcal{R}^2}{8} \right) +\mathcal{W}(\phi) \right]\,,
\end{equation}
where $\mathcal{R}_{ab}$ denotes Ricci curvature on the boundary $\partial M$. The radius of AdS in our units is $l=1/2$. The superpotential $\mathcal{W}$ appearing in \eqref{canon} satisfies this relation:
\begin{equation} \label{fundrelationpotential}
V = \frac12 \left(-\frac32 \mathcal{W}^2 + g^{ij} \partial_i \mathcal{W} \partial_j \mathcal{W} \right)\,.
\end{equation}
For our purposes, it is sufficient to know the form of $\mathcal{W}$ close to the AdS vacuum. More precisely, we can write the scalar potential in terms of the fields $\phi_{m^2}$ which (1) have canonical kinetic terms in a neighborhood of the minimum of the potential, and (2) diagonalize the mass matrix. In terms of these fields, the superpotential is
\begin{equation}
\mathcal{W}_{\pm} = 4 + a_{\pm} \phi_{-2,a}^2 + b_{\pm} \phi_{-2,b}^2 + c_{\pm} \phi_{18}^2 + d_{\pm} \phi_{10}^2 + e_{\pm} \phi_{4}^2 + \order{\phi^3}  \ .
\end{equation}
The modes with $m^2 l^2 = 4, 10, 18$ fall off faster than $1/\sqrt{g_3}$ near the boundary of AdS and hence do not contribute to \eqref{canon}. For the light modes with $m^2 l^2 = -2$, the coefficients  $a_{\pm}$ and $b_\pm$ are the conformal dimensions of the operators dual to these scalar fields \cite{deBoer:1999tgo}. Each of them can be $1$ or $2$. The divergences in the action cancel if we use $a = b = 1$ (see for example the discussion in \cite{Papadimitriou:2006dr}). Therefore, it is sufficient to take into account
\begin{equation}
\mathcal{W}= \mathcal{W}_{-}= 4 + \phi_{-2,a}^2 + \phi_{-2,b}^2 + \ldots
\end{equation}
Finally, following the prescription of \cite{Bianchi:2001kw,Martelli:2002sp,Korovin2014}, the presence of the massive vector field requires the presence of an additional counterterm of the form
\begin{equation} \label{count_yegor}
I_{ct,A} \propto \int_{\partial M} d^3x \sqrt{g_3}  \,  \mathcal{B}_{\mu} \mathcal{B}^{\mu}\,.
\end{equation}
However, in our subspace of solutions this counterterm does not give any finite contribution to the renormalized action, as one can see from the asymptotic expansion of $\zeta$ in \eqref{exp_infinity}.

\subsection{Electric solution}\label{el_onshell}

We now compute the on-shell value of the renormalized action for purely electric black holes, following \cite{Gauntlett:2009bh,Batrachenko:2004fd}. The magnetic ones follow along the same lines. 

For purely electric and purely magnetic configurations, the terms of the form $F \wedge F$ in the action \eqref{N2action} vanish. Substituting the trace of the Einstein equation \eqref{useful_einst} into the action \eqref{N2action}, we have
\begin{align}
I &= \int{\td^4 x \sqrt{g} \left( \frac{1}{2} \im \mathcal{N}_{\Lambda \Sigma} F_{\mu \nu}^{\Lambda} F^{\mu \nu \Sigma} + V \right)}\,, 
\end{align}
and  making use of the $tt$ component of the Einstein's equations \eqref{useful_einst} in Appendix, we obtain
\begin{align}
I &= \int{\td^4x \sqrt{g} \left( R_t^t + 2 \im \mathcal{N}_{\Lambda \Sigma} F_{tr}^{\Lambda} F^{tr \Sigma} - 2 A_{t}^{\Lambda} A^{t, \Sigma} k_{\Lambda}^u k_{\Sigma\, u}  \right)} \ . \label{manip1}
\end{align}
Remarkably, the quantity $R_t^t$ can be written as a total derivative \cite{Batrachenko:2004fd}
\begin{equation}
R_t^t = \frac{1}{\sqrt{g}} \frac{d}{dr} (\sqrt{g_3} \, \,\Theta_{t}^t) \ .
\end{equation}
Now, the term  $2 \im \mathcal{N}_{\Lambda \Sigma} F_{tr}^{\Lambda} F^{tr \Sigma} $ can be written in the basis  spanned by the two massless vectors  $\mathcal{A}_i$  and the  massive one $\mathcal{B}$, in the form $G_{\mathcal{A}_i} F_{\mathcal{A}_i} + G_{\mathcal{B}} F_{\mathcal{B}}$. Recalling the definitions  \eqref{chargesflux} and \eqref{Fdual} and making use of Maxwell's equations \eqref{maxw_e} for the  massless vectors, we have
\be
\partial_{r} (\sqrt{g} \,\, G_{\mathcal{A}_i}^{tr} )   =0\,, \qquad \sqrt{g} \,\, G_{\mathcal{A}_i}^{tr}  = Q_i \,,
\ee
hence
\be
\int \td r \sqrt{g} \,\,G_{\mathcal{A}_i} F_{\mathcal{A}_i}  = \int  \td r \, \left[ Q_i \mathcal{A}_i \right]' \,.
\ee
For the massive vector we again use the Maxwell's equation and we take into account the contribution of the third term in \eqref{manip1}. After integration by parts, and recalling the boundary conditions for the massive vector (which vanishes at the event horizon and asymptotically), the on-shell action  \eqref{manip1} assumes the following form
\begin{align} \label{onshell}
I &= 8 \pi \bar{\beta} \int dr \left[ \sqrt{g_3} \,\, \frac{\Theta_t^t}{2}+  Q_{i} \mathcal{A}_{t,i} \right]' \nonumber \\
&= 8 \pi \bar{\beta}  \big( \sqrt{g_3} \,\, \frac{\Theta_t^t}{2}  + Q_i  \mathcal{A}_{t,i} \big)\bigg|_{r_H}^{r_C} \,,
\end{align} 
where we have regulated the action using a radial cutoff $r_C$ which will be sent to infinity after the integration. The factor $\bar{\beta}=2\pi/T$ comes from the integration over the Euclidean time direction. Notice that the expression \eqref{onshell} gives contributions both at the horizon, located at $r =r_H$ and at the boundary.

\noindent Adding the counterterms to this action, we have
\begin{equation}
I_{GH} =  (4 \pi \bar\beta) \frac12 \,  r_C  \, e^{-\beta(r_C)/2} \left[ r h'+h \left(4-r \beta '\right)\right]_{r_C}\,.
\end{equation}
Moreover, using $\mathcal{R} = 2/r^2$ and $\mathcal{R}_{ab} \mathcal{R}^{ab} = 2 / r^4$, the counterterms action \eqref{canon} becomes
\begin{equation}
I_{ct}  = \frac12 (4 \pi \bar\beta) \left. e^{-\beta/2} \sqrt{h} \, [1+2 r^2 ( 4 + \phi_{-2,a}^2 + \phi_{-2,b}^2 + \ldots)] \right|_{r_C}\,.
\end{equation}

The complete renormalized on-shell action $I_{ren}$ \eqref{ren_os_action} can thus be calculated using the asymptotic and horizon expansions of the fields \eqref{exp_infinity} and \eqref{exp_horizon}. We obtain 
\begin{equation}
\frac{I_{ren}}{4 \pi \bar\beta}= -\frac{1}{2}  (h_{(3)} -4  e_{a,2} \,e_{a,1}- 4e_{b,2}\, e_{b,1}) - \frac14 [ r^2 e^{-\beta/2} h']_{r_H} - \phi^{\mathcal{A}i} Q_{i}\ , \label{eq:actionFreeEnergy}
\end{equation}
where $e_{a,1},e_{b,1}$ are the leading falloff at the boundary of the light scalar modes ($m^2 l^2=-2$) as defined in \eqref{as_fall2}\footnote{The values of the mass eigenstate coefficients $e_{a,i}$, $e_{b,i}$ are related to the expansion parameters we used in \eqref{eq:scalarExpansion} as follows:
\begin{align}
e_{a,1} &= - \sqrt3 b_{(1)}\,, \qquad e_{a,2}= -\sqrt3 \,b_{(2)} \ , & e_{b,1} &= -\sqrt3 \tau_{(1)} \qquad e_{b,2}= -\frac{\sqrt3(3b_{(1)}^2+2(\tau_{(1)}^2+5\tau_{(2)}))}{10}\,. 
\end{align}
}. 

The right-hand side of \eqref{eq:actionFreeEnergy} is the expression for the free energy, as anticipated. The term evaluated at $r_H$ is nothing but $T S$, using \eqref{temperature_expl},\eqref{BHentropy}. Furthermore, the first term in bracket gives the black hole mass  \eqref{mass_eig}, hence
\be \label{Gibbs}
\frac{I_{ren}}{4 \pi \bar\beta}= M -TS - \phi^{\mathcal{A}_i} Q_i\,.
\ee

This expression for the mass agrees with the one obtained from the renormalized boundary stress-energy tensor $\tau^{ab}$ \cite{Brown:1992br}
\begin{align}
  M &= Q_t = \frac{1}{16 \pi} \int_{\Sigma} \sqrt{\sigma} \, u_a \tau^{ab} \xi_t \ , & \tau^{ab} &= \frac{2}{\sqrt{g_3}} \frac{\delta I}{ \delta g_{3,ab}} \ ,
\end{align}
where $\xi^{a} \partial_{a}= \partial_t$ is the Killing vector of the time translation isometry of the boundary metric $g_{3 ab}$, $\Sigma$ is a constant time slice on the boundary  $\partial M$ with induced metric $\sigma$, and $u^a$ is the timelike unit normal vector to $\Sigma$ on $\partial M$ (see for instance \cite{Caldarelli:1999xj}). We have:
\begin{equation}
\tau^{tt} = -(\Theta^{ab}-\Theta g_3^{ab}) +\mathcal{W} g_3^{ab} - l \left( \mathcal{R}^{ab} -\frac12 g_3^{ab} \mathcal{R}  \right)
\end{equation}
which yields exactly \eqref{mass_eig}.

We have seen then that the choice of counterterms \eqref{count_GH} \eqref{canon} reproduces the Gibbs free energy \eqref{Gibbs}. From \cite{Gauntlett:2009bh} one can see that $I_{ren}$ is stationary for fixed temperature and chemical potential, and in particular for fixed $e_{a,1}$ and $e_{b,1}$, as is the case in our solutions\footnote{It is nevertheless possible to impose different boundary conditions (Neumann, mixed) for the scalar fields with $m^2=-2$, by choosing appropriate boundary counterterms. For simplicity we restrict here to the case of fixed $e_{a,1}=e_{b,1}=0$ fixed, but it would be interesting to analyze the other cases as well.}.

Finally, let us mention that if we are instead interested in the canonical ensemble, we need to add to the action $I_{ren}$ obtained before the additional finite counterterm
\begin{equation}
I_{HR} = -\int d^3x \sqrt{g_3} \, \,  \im \mathcal{N}_{\Lambda \Sigma} \, n_\mu F^{\mu \nu,\Lambda} A_{\nu}^{\Sigma} \,,
\end{equation}
called Hawking-Ross counterterm  \cite{Hawking:1995ap}. The total action $I_{ren}+ I_{HR}$ is then stationary for fixed electric \emph{charges}, hence the first law reads
\begin{equation} \label{first_law}
\td M = T \td S + \phi^{\mathcal{A}_i} \td Q_i\,,
\end{equation}
with Helmholtz free energy 
\begin{equation} \label{helm}
F_{Helmholtz}  =M - TS\,.
\end{equation}

\bigskip

The on shell action for the purely magnetic configuration can be worked out analogously, for instance along the lines of \cite{Gnecchi:2014cqa,Gnecchi:2016auh}.

We have tested the accuracy of our numerics by verifying that the first law \eqref{first_law} is satisfied for infinitesimal changes of the conserved quantities $\delta M$, $\delta S$, $\delta Q_{i}$. We have moreover computed the renormalized on shell action by numerically integrating \eqref{onshell} and found agreement with the expression we obtained in eq.\eqref{Gibbs}.

%% file: PhaseTransitions.tex
In what follows we analyze the thermodynamics of the novel black hole solutions in the canonical ensemble, namely for fixed values of temperature and electromagnetic charges. Moreover, in what follows, we restrict to configurations with boundary conditions $e_{a,1} =e_{b,1}=0$ for the scalar fields of mass $m^2 l^2=-2$. We consider the thermodynamic analysis of the purely electric configuration, since (as opposed to the purely magnetic ones) these have a nontrivial profile for the massive vector field. For simplicity, we will moreover restrict in our discussion to solutions with $Q_1=0$: allowing for a nonzero value of the other charge $Q_2$ is sufficient for the solution to support a nontrivial massive vector profile\footnote{Setting $P^2 = Q_2 = 0$ (and keeping $e_{a,1} = e_{b,1} = 0$) only yields the AdS-Reissner-Nordstr\"om solution.}

We compare configurations with fixed electric charge $Q_2$, but with different values of the radius of the event horizon (hence different values for the entropy and mass of the black hole). Whenever there are multiple solutions with the same $T$ and $Q_2$, those which minimize the Helmholtz free energy \eqref{helm} will dominate the thermodynamic ensemble. 

\begin{figure}[htbp]
\centering
{\includegraphics[width=\textwidth]{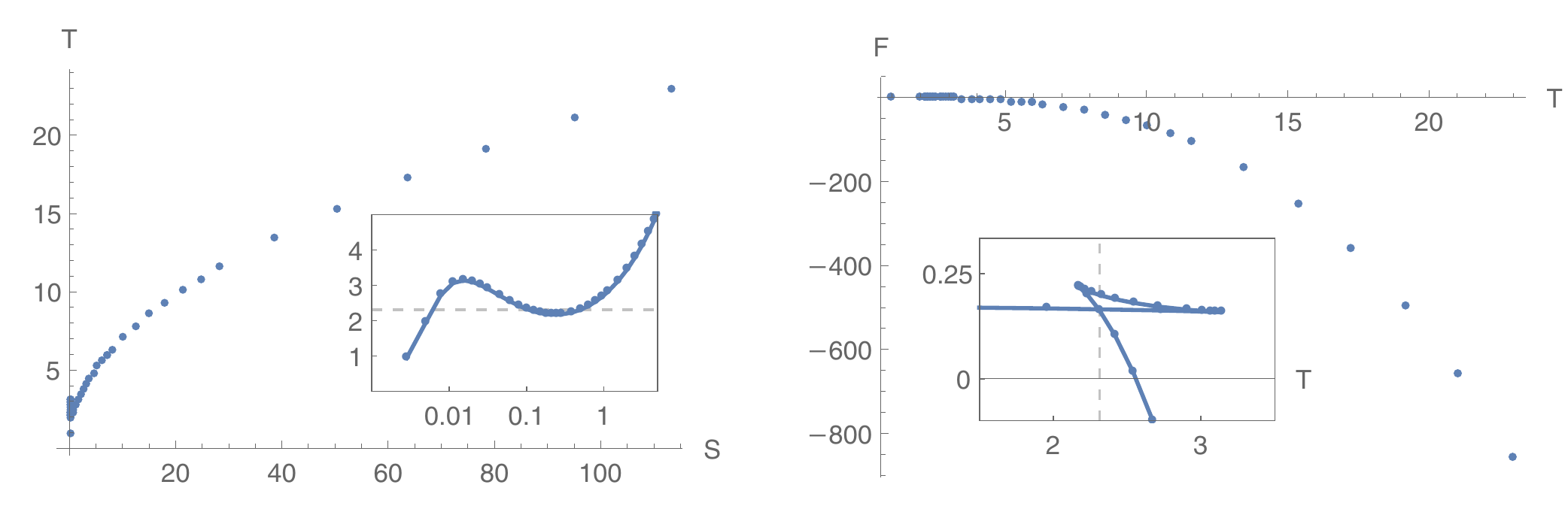}}
\caption{Plot of the temperature in function of the black hole entropy for the set of solutions with $Q_2 = 0.15$.  At the temperature $T \approx 2.4$ the derivative of the free energy exhibits a discontinuity, revealing a first order phase transition. Notice that the horizontal axis in the first plot is logarithmic, hence the Maxwell area law is not explicitly visible. \label{fig:ThD}}
\end{figure}

We were able to find families of black holes by sampling the space of solutions with different discrete values of the event horizon. The results of this procedure are plots like those in Figure \ref{fig:ThD}. It turns out that for values $|Q_2| \geq Q_c$ where $Q_c \approx 0.17$, the temperature plotted in function of the black hole entropy is a monotonically increasing function, while if we lower the charge to values $|Q_2| < Q_c$, for a suitable temperature range we are able to find three branches of solutions, characterized by three possible different values of the entropy. We call them small, medium and large black holes, where the size relates to the black hole radius as compared with the AdS radius. In our conventions, this happens for black hole entropies of order 1, see Figure \ref{fig:ThD}.

With reference to the same figure, in the right panel we plot the free energy as a function of the temperature, for the same set of solutions. It is clear that for $Q_2 < Q_c$ the free energy exhibits a discontinuity in the first derivative for a value of temperature $T \approx 2.4$ (see Figure \ref{fig:ThD}). This signals the onset of a small-large black hole first order phase transition, in all similarity with the phase transition for Reissner-Nordstr\"om in AdS space-time  found in the seminal papers of \cite{Chamblin:1999hg,Chamblin:1999tk} and \cite{Caldarelli:1999xj}. The phase transitions become a crossover for charges $Q_2 > Q_c$, while second order for the critical charge $Q_2 \approx 0.17$. Notice that, despite appearing almost horizontal in the plot, the free energy for each branch is always monotonically decreasing, as it should be since $\partial F /\partial T = -S $. Lastly, the medium-sized black holes are always thermodynamically disfavored since their free energy is always greater than that of the other two black hole branches. They also have negative specific heat 
 \begin{equation}
 C_S = T \left(\frac{\partial S}{ \partial T} \right)_{Q,T}
 \end{equation}
while the small and large have positive specific heat. 

\begin{figure}[htbp]
\centering
{\includegraphics[width=0.5\textwidth]{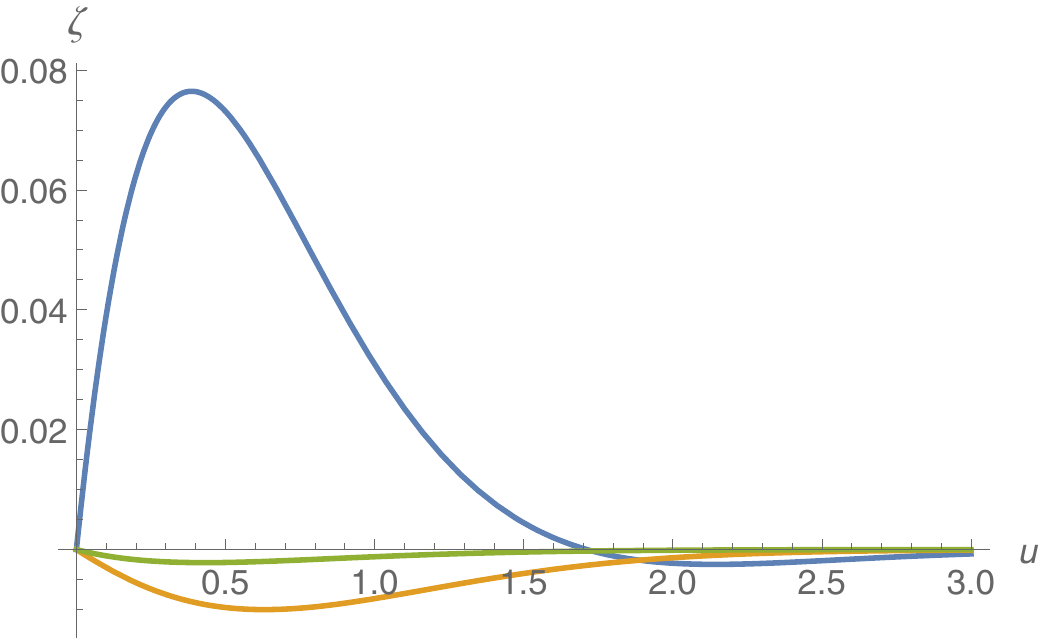}}
\caption{Plot of the radial profile for the massive vector field for small black holes (blue, $r_H = 0.03$), medium ones (orange, $r_H = 0.07$) and large ones (green, $r_H = 0.14$).
\label{fig:massiveVector}}
\end{figure}

It is instructive to plot the behavior of the massive vector field for the three different black hole branches, as done in Figure \ref{fig:massiveVector}. We notice that, with reference to the black hole family in figure \ref{fig:ThD}, the small black holes (blue line in the plot) have a profile for the massive vector with two extrema: one at a positive value close to the black hole horizon and a smaller peak at a negative value somewhat further away. The first peak goes away as the size of the black hole increases. The medium black holes (orange in Figure \ref{fig:massiveVector}) have only a minimum in the $\zeta$-profile, which is however more pronounced and closer to the black hole horizon (in terms of $u$, i.e. with $r_H$ scaled out). For large black holes (in green), the minimum in the massive vector profile becomes ever less pronounced. It settles at $u_\text{extr} \approx 0.34$, which corresponds to $r_\text{extr} \approx 1.4 r_H$. To sum up, during the phase transition from small to large black holes, the radial coordinate $r_\text{extr}$ corresponding to the maximum value of the massive vector field  increases, moreover the massive vector field considerably decreases in absolute value. The matter outside the horizon gets "swallowed" into the black hole as the small-large phase transition happens.

\begin{figure}[htbp]
\centering
{\includegraphics[width=0.45\textwidth]{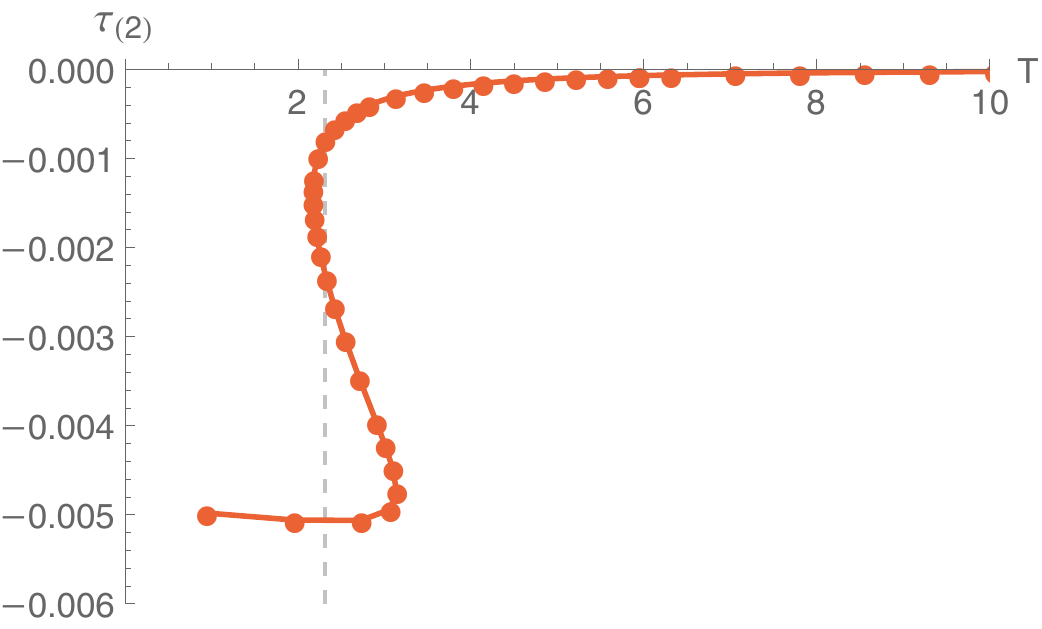}}
{\includegraphics[width=0.45\textwidth]{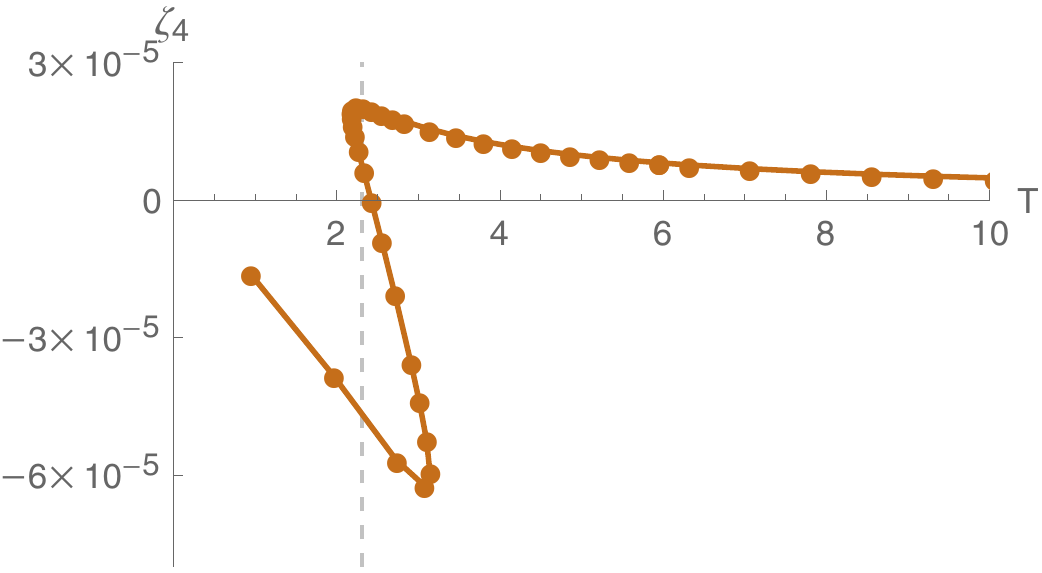}}
\caption{On the left, we plot of the expectation value of the operator of dimension 2 dual to the light scalar mode with mass $m^2 l^2=-2$ in function of the temperature of the black hole. On the right the plot of the parameter $\zeta_4$ in function of the temperature. \label{fig:tau2zeta4}}
\end{figure}

We now analyze the scalar field asymptotic expansion. The mode $e_{b,1}$ corresponds to the expectation value of an operator of conformal dimension 2, $\braket{\mathcal{O}_2} = -\tau_{(2)}/2$. This is the order parameter of our phase transition. The value of $\tau_{(2)}$ in function of the temperature is visualized in Figure \ref{fig:tau2zeta4}. We see that its absolute value decreases during the phase transition from low to high temperature. Moreover, its behavior resembles that of the isotherms for the Van der Waals system (liquid/gas -like phase transition). This is reminiscent of what happens for black holes solutions of Fayet-Iliopoulos gauged supergravity \cite{Hristov:2013sya,Toldo:2016nia}.

The interpretation of the value of $\zeta_4$ is more subtle: due to interaction terms with the light scalar fields, the term proportional to $\zeta_4$ does not dominate its asymptotic expansion \eqref{eq:vectorExpansion}. Its interpretation as the expectation value of the corresponding operator with $\Delta=5$ needs verification, by means of the identification of the correct renormalized conjugate momenta of $\zeta$, see for instance \cite{Bianchi:2001kw,Papadimitriou:2010as}. We nevertheless provide the behavior of $\zeta_4$ in the second graph of Figure \ref{fig:tau2zeta4}, where we can see once again that the small-large black hole phase transition manifest itself as a decrease in the absolute value of this parameter. 

We conclude by highlighting yet another difference with respect to the holographic superconductor phase transition. The process we have described here for the new class of solutions involves two phases where the condensate is never vanishing, namely the massive vector field is always switched on. There is no restoring of the broken symmetry for a finite temperature, as opposed to \cite{Gubser:2008px,Hartnoll:2008vx,Hartnoll:2008kx,Denef:2009tp}, where the preferred phase for high temperatures is the scalarless Reissner-Nordstr\"om solution, with no scalar condensate.

%% file: Outlook.tex
In this work we have constructed novel numerical solutions of $\mathcal{N}=2$ gauged supergravity coupled to vector and hypermultiplets. This four-dimensional theory arise as consistent truncation of M-theory on the manifold $M^{111}$ and it is endowed with one Betti vector multiplet. The presence of the latter, corresponding to light degrees of freedom (two scalars of mass $m^2=-2$ and one massless vector), allows for the construction of black hole solutions with non vanishing massive vectors. This fact was noticed in the BPS case as well \cite{Halmagyi:2013sla} and it would be interesting to understand its deeper origin, in relation to brane world volume gauge theories.

We have moreover analyzed the thermodynamics of the black hole configurations, revealing two branches of stable solutions: the so-called small black holes and the large ones. A small-large black hole phase transition was found, during which the massive vector field decreases in absolute value, being "swallowed" inside the black hole.

The black holes constructed here serve as the starting point for future analysis of bound states in AdS space-times, in view of applications to glassy systems \cite{Anninos:2011vn,Anninos:2012gk,Anninos:2013mfa}. The next step in this direction will be to establish the possible existence of finite temperature bound states composed of the black holes backgrounds which we have discovered, surrounded by smaller probe black holes. One of the differences with respect to the analysis carried out in \cite{Anninos:2011vn,Anninos:2012gk,Anninos:2013mfa} is that in the present case there are additional interactions between the probes and the massive vector field condensate. We expect this new feature to play a role in the equilibrium condition for the charged probes. Let us mention that a similar stability analysis for a smaller truncation of the $M^{111}$ and $Q^{111}$ models has been carried out in \cite{Klebanov:2010tj}, where the stability of probe M2-branes in the background of an AdS black brane was studied. Our analysis will be closely related to this latter, albeit the background assumes a more elaborate form.

Moreover, as mentioned in the introduction, the systems of bound states with magnetic charges would come with strings attached, due to Meissner effect. It would be interesting to quantify the effect of such strings stretched from the horizon to the probes, and among the probes themselves. The solutions would manifest themselves in the lower dimensional supergravity system as vortex-like solutions, like those constructed in \cite{Gregory:2014uca} on AdS black holes. This is another point that should be taken into account.

We will then be able to chart the parameter space of allowed stable and metastable configurations. Subsequently, the relaxation dynamics of such bound systems can be studied, and one can verify if they exhibit logarithmic aging behavior which is typical of many amorphous systems. The overall picture emerging from \cite{Anninos:2011vn,Anninos:2012gk,Anninos:2013mfa} was that, upon cooling (decreasing the temperature), a liquid single-centered horizon corresponding to the liquid phase of matter, can turn into a fragmented, disordered one corresponding to a glassy phase. To make this analogy more precise, it would be interesting to compute holographic transport coefficients for the composite systems, such as shear viscosity and conductivity. We hope to report back on these points in the near future.

%% file: Conventions.tex
We adopt signature $[-,+,+,+]$ and Riemann-Christoffel tensor and the Ricci tensor are defined as
\begin{equation}
{R^{\rho}}_{\sigma \mu \nu} = -(\partial_{\mu} \Gamma^{\rho}_{\nu \sigma}- \partial_{\mu} \Gamma^{\rho}_{\nu \sigma} + \Gamma^{\rho}_{\mu \lambda}\Gamma^{\lambda}_{\nu \sigma} - \Gamma^{\rho}_{\nu \lambda}\Gamma^{\lambda}_{\mu \sigma})\,, \qquad
{R^{\rho}}_{\sigma \rho \nu} = R_{\sigma \nu}\,.
\end{equation}
The Einstein's equation then read:
\begin{equation}
-(R_{ \mu \nu}- \frac12 R g_{\mu \nu})=  T_{\mu \nu}\,,
\end{equation}
where $T_{00}$ is negative. We furthermore take $c = G_{N} =1$. The Levi-Civita symbol $\epsilon$ is defined as in \cite{Carroll:1997ar} 
\begin{align}
\epsilon_{0123} &= 1 = - \epsilon^{0123}  \ , & \epsilon^{\mu \nu \rho \sigma} &= \sqrt{-det\, g} \ {{e}^{\mu}}_{a} {{e}^{\nu}}_b {{e}^{\rho}}_c {{e}^{\sigma}}_d \epsilon^{abcd} \ .
\end{align}

We work in a symplectic gauge where the covariantly holomorphic sections are $X^{\Lambda} =(1,t_1^2,t_2^2)$. With these conventions the K\"ahler potential reads
\begin{equation}
\mathcal{K} = - \log [(t_1 + \bar{t}_1)^2 (t_2+\bar{t}_2)]\,,
\end{equation}
which gives the following components for the metric of the vector multiplet scalars:
\begin{equation}
g_{1\bar{1}} = \frac{2}{(t_1+\bar{t}_1)^2} \,, \qquad g_{2\bar{2}} = \frac{1}{(t_2+\bar{t}_2)^2}\,.
\end{equation}

The components of the period matrix $\mathcal{N}$ are:
\begin{align}
\im\mathcal{N}_{00} &= -\frac{\tau_2 \left(b_1^2+\tau_1^2\right)^2 \left(2 b_1^2 \tau_2^2+\tau_1^2 \left(b_2^2+\tau_2^2\right)\right)}{\tau_2^2 \left(2 b_1^2+\tau_1^2\right)^2+b_2^2 \tau_1^4} \ , &  \re\mathcal{N}_{00} &= -\frac{2 b_1^2 b_2 \tau_2^2 \left(b_1^2+\tau_1^2\right)^2}{\tau_2^2 \left(2 b_1^2+\tau_1^2\right)^2+b_2^2 \tau_1^4} , \nonumber \\
\im\mathcal{N}_{10} &= \frac{2 b_1 b_2 \tau_1^2 \tau_2 \left(b_1^2+\tau_1^2\right)}{\tau_2^2 \left(2 b_1^2+\tau_1^2\right)^2+b_2^2 \tau_1^4}  \ , &  \re\mathcal{N}_{10} &= \frac{2 b_1 \tau_2^2 \left(b_1^2+\tau_1^2\right) \left(2 b_1^2+\tau_1^2\right)}{\tau_2^2 \left(2 b_1^2+\tau_1^2\right)^2+b_2^2 \tau_1^4}  \ , \displaybreak[0] \nonumber \\
\im\mathcal{N}_{20} &= \frac{b_1^2 \tau_2 \left(\tau_2^2 \left(2 b_1^2+\tau_1^2\right)-b_2^2 \tau_1^2\right)}{\tau_2^2 \left(2 b_1^2+\tau_1^2\right)^2+b_2^2 \tau_1^4}  \ , & \re\mathcal{N}_{20} &= \frac{b_2 \tau_2^2 \left(2 b_1^4+2 b_1^2 \tau_1^2+\tau_1^4\right) + b_2^3 \tau_1^4}{\tau_2^2 \left(2 b_1^2+\tau_1^2\right)^2+b_2^2 \tau_1^4}  \ , \displaybreak[0] \nonumber \\
\im\mathcal{N}_{11} &= -\frac{2 \tau_1^2 \tau_2 \left(2 b_1^2+\tau_1^2\right)}{\tau_2^2 \left(2 b_1^2+\tau_1^2\right)^2+b_2^2 \tau_1^4} \ , & \re\mathcal{N}_{11} &= \frac{2 b_2 \tau_1^4}{\tau_2^2 \left(2 b_1^2+\tau_1^2\right)^2+b_2^2 \tau_1^4} \ , \displaybreak[0] \nonumber \\
\im\mathcal{N}_{12} &= \frac{2 b_1 b_2 \tau_1^2 \tau_2}{\tau_2^2 \left(2 b_1^2+\tau_1^2\right)^2+b_2^2 \tau_1^4}  \ , & \re\mathcal{N}_{12} &= \frac{2 b_1 \tau_2^2 \left(2 b_1^2+\tau_1^2\right)}{\tau_2^2 \left(2 b_1^2+\tau_1^2\right)^2+b_2^2 \tau_1^4}  \ , \nonumber \\
\im\mathcal{N}_{22} &= -\frac{\tau_2 \left(2 b_1^2 \tau_2^2+\tau_1^2 \left(b_2^2+\tau_2^2\right)\right)}{\tau_2^2 \left(2 b_1^2+\tau_1^2\right)^2+b_2^2 \tau_1^4} \ , & \re\mathcal{N}_{22} &= -\frac{2 b_1^2 b_2 \tau_2^2}{\tau_2^2 \left(2 b_1^2+\tau_1^2\right)^2+b_2^2 \tau_1^4}  \ . \label{eq:cN}
\end{align}

%% file: EoM.tex
Let us now recapitulate how we arrived to the ansatz \eqref{metricR} - \eqref{ans_vec}. We are looking for static spherically symmetric configurations,  which are characterized by the space-time Killing vectors
\begin{align} \label{killingit}
&\partial_t \ , & &\partial_{\varphi} \ , & &\cos \varphi \ , & &\partial_{\theta} - \tan \theta \, \sin \varphi \, \partial_{\varphi} \ , & &\sin \varphi \, \partial_{\theta} + \cot \theta \, \cos \varphi \, \partial_{\varphi} \,.
\end{align}
This justifies the ansatz \eqref{metricR}, as it is the most general static and spherically symmetric metric. Moreover, the scalar fields can only depend on the radial coordinate $r$. As explained in detail in \cite{Chimento:2015rra}, the requirement of the field strengths to be invariant under the variations generated by \eqref{killingit} leads to the following nonvanishing components for the field strengths
\begin{equation}
F_{tr}^{\Lambda}=f^{\Lambda} (r)\,, \qquad  \qquad F_{\theta \phi }^{\Lambda}  = P^{\Lambda}(r) \sin \theta\,, \qquad \Lambda =0,1,2\,.
\end{equation}
Moreover, the Bianchi identities $\partial_{[ \mu} F^{\Lambda}_{\nu \rho ]} = 0$ imply that the functions $P^{\Lambda}$ must be constant. Therefore we end up with the form \eqref{ans_vec} for the vector fields presented in Section \ref{static_BH_ansatz} .

Given this ansatz, we are now ready to analyze Maxwell's equations, which will further constrain the black hole charges given the presence of a nontrivial source term. Maxwell's equation reads
\begin{equation}  \label{maxwell_equation}
\epsilon^{\mu \nu \rho \sigma} D_{\nu} ( G_{\rho \sigma, \Lambda}) + \frac12 k_{\Lambda \, u} D^{\mu} q^u=0\,.
\end{equation}
Given the form of the gauging \eqref{kv} and the radial dependence of the hyperscalars (with the field $a$ consistently set to zero), the $r$ and $\theta$ components of Maxwell's equations are automatically satisfied. The $\varphi$ component is
 \begin{equation} \label{constraint0}
  k_{\Sigma \,u} k_{\Lambda}^u(q) P^{\Lambda} = 0 \,,
\end{equation}
which, given \eqref{eigen_vec} and \eqref{ans_vec}, is satisfied if $P^m =0$. In other words, we have found that the $\varphi$ component of the massive vector field must vanish. This leaves just the other two conserved quantities $P^i$, $i=1,2$ denoting the magnetic charges of the unbroken $U(1)s$.
 
The other nontrivial component of Maxwell's equation is in the $t$ direction. Defining radially dependent functions $e_{\Lambda}(r)$ such that
 \begin{equation}
 F^{\Lambda}_{tr} =\frac{e^{-\beta/2}}{r^2} \im (\mathcal{N}^{-1})^{\Lambda \Sigma} (e_{\Sigma}(r)-\re \mathcal{N}_{\Sigma \Gamma} P^{\Gamma}),
 \end{equation}
 the $t$ component of the Maxwell's equation is
\begin{equation} \label{maxw_e}
 \partial_{r} \,  e_{\Lambda} (r)= \frac{e^{\beta/2}}{h} A^{\Gamma}_t \, k_{\Gamma}^u \, k_{\Lambda\,u} \,.
\end{equation}
The right-hand side is only nontrivial for a certain combination of the $e_\Lambda$, namely the one corresponding to $\mathcal{B}$ in \eqref{eigen_vec}. For the other two combinations, which correspond to the massless gauge fields, there is no source on the right-hand side. These linear combinations of $e_{\Sigma}$ are then constant and correspond to the two conserved electric charges. 

\bigskip

The scalar equations of motion are as follows (we avoided inserting the explicit form of the field strengths, as their expression quickly becomes very cumbersome). \\
scalar $\ \tau_1$)
\begin{align} \label{scalar_tau1}
\frac{e^{\beta/2}}{r^2} \partial_{r} \left( e^{-\beta/2}r^2 h \frac{\partial_r \tau_1}{\tau_1^2} \right) &\eq -h \frac{ (\partial_r b_1)^2}{\tau_1^3} +  \frac{ \partial \im\mathcal{N}_{\Lambda \Sigma}} {\partial \tau_1} F_{\mu \nu}^{\Lambda} F^{\mu \nu \Sigma} +  \frac12 \frac{\partial \re\mathcal{N}_{\Lambda \Sigma} }{\partial \tau_1} \epsilon^{\mu \nu \rho \sigma} F_{\mu \nu}^{\Lambda} F_{\rho \sigma}^{\Sigma} \nonumber  \\
&\pheq + 2 \sigma^4  \frac{\tau_1}{\tau_2} +16 \frac{\sigma^2 }{\tau_1^2} -\frac{ \sigma^4 [(4 b_1 b_2 + 2 b_1^2 +6)^2+16b_1^2 \tau_2^2 ]}{2\tau_1^3 \tau_2} \ ,
\end{align}
scalar $\tau_2$)
\begin{align}\label{scalar_tau2}
\frac{e^{\beta/2}}{r^2} \partial_{r} \left( e^{-\beta/2}r^2 h \frac{\partial_r \tau_2}{\tau_2^2}  \right) &\eq -h \frac{ (\partial_r b_2)^2}{\tau_2^3} + \frac{ \partial \im\mathcal{N}_{\Lambda \Sigma}} {\partial \tau_2} F_{\mu \nu}^{\Lambda} F^{\mu \nu \Sigma} + \frac12  \frac{\partial \re\mathcal{N}_{\Lambda \Sigma} }{\partial \tau_2} \epsilon^{\mu \nu \rho \sigma} F_{\mu \nu}^{\Lambda} F_{\rho \sigma}^{\Sigma} \nonumber \\
&\pheq - 2\frac{\sigma^4 \tau_1^2}{\tau_2^2}+4\sigma^4 + 16 \frac{\sigma^2 }{\tau_2^2}+8 \frac{b_1^2}{\tau_1^2} \nonumber \\
&\pheq - \frac{ \sigma^4 [(4 b_1 b_2 + 2 b_1^2 +6)^2+4(\tau_1^2(b_1+b_2)^2 +4b_1^2 \tau_2^2)]}{2\tau_1^2 \tau_2^2} \ ,
\end{align}
scalar $b_1$)
\begin{align} \label{scalar_b1}
\frac{e^{\beta/2}}{r^2} \partial_r \left( \frac{ r^2 e^{-\beta/2} h }{\tau^2} \partial_r b_1 \right) &\eq \frac{ \partial \im\mathcal{N}_{\Lambda \Sigma}} {\partial b_1} F_{\mu \nu}^{\Lambda} F^{\mu \nu \Sigma} + \frac12  \frac{\partial \re\mathcal{N}_{\Lambda \Sigma} }{\partial b_1} \epsilon^{\mu \nu \rho \sigma} F_{\mu \nu}^{\Lambda} F_{\rho \sigma}^{\Sigma} \nonumber \\
&\pheq + \frac{ \sigma^4\left( 2  (4 b_1 b_2 +2 b_1^2 +6) (4b_2+ 4b_1) \right)}{4 \tau_1^2 \tau_2} \nonumber \\
&\pheq + \frac{ \sigma^4\left( 4 (4 \tau_1^2 (b_1+b_2) b_1 +8 \tau_2^2 b_1)\right)}{4 \tau_1^2 \tau_2} \ ,
\end{align}
scalar $b_2$)
\begin{align}\label{scalar_b2}
\frac{e^{\beta/2}}{r^2} \partial_r \left( \frac{ r^2 e^{-\beta/2} h }{\tau^2} \partial_r b_2 \right) &\eq \frac{ \partial \im\mathcal{N}_{\Lambda \Sigma}} {\partial b_2} F_{\mu \nu}^{\Lambda} F^{\mu \nu \Sigma} + \frac12 \frac{\partial \re\mathcal{N}_{\Lambda \Sigma} }{\partial b_2} \epsilon^{\mu \nu \rho \sigma} F_{\mu \nu}^{\Lambda} F_{\rho \sigma}^{\Sigma} \nonumber \\
&\pheq + \frac{2 \sigma^4\left( 2  (4 b_1 b_2 +2 b_1^2 +6) (4b_1) +4 (4 \tau_1^2 (b_1+b_2) b_2)\right)}{4 \tau_1^2 \tau_2}  \ ,
\end{align}
scalar $\sigma$)
\begin{align} \label{scalar_sigma}
\frac{e^{\beta/2}}{r^2} \partial_r \left(  \frac{e^{-\beta/2} r^2 h}{\sigma^2} \partial_r \sigma \right) &\eq -24 \sigma^3 \zeta^2 \frac{e^{\beta}}{h} + 2\sigma^3 \left(\frac{\tau_1^2}{\tau_2}+2\tau_2\right)-8 \sigma \left(\frac{2}{\tau_1}+\frac{1}{\tau_2}\right) \nonumber  \\
&\pheq + \frac{ 4\sigma^3 \left(4 b_1 b_2+2 b_1^2+6 \right)^2+16\sigma^3 \left(2\tau_1^2 (b_1+b_2)^2+4 b_1^2\tau_2^2\right)}{8\tau_1^2\tau_2} \ .
\end{align}
The Einstein equations are:
\begin{align}
-(R_{\mu \nu} - \frac12 g_{\mu \nu} R) &\eq g_{\mu \nu} g_{i \bar{\jmath}} \partial^{\sigma}t^{i} \partial_{\sigma} \bar{t}^{\bar{\jmath}} -2  g_{i \bar{\jmath}} \partial_{\mu}t^{i} \partial_{\nu}\bar{t}^{\bar{\jmath}} + g_{\mu \nu}  h_{uv} D^{\sigma}q^{u} D_{\sigma} q^{v} \nonumber \\
&\pheq - 2  h_{uv} D_{\mu}q^u D_{\nu} q^v - \frac12 I_{\Lambda \Sigma} g_{\mu \nu} F_{\rho \sigma}^{\Lambda} F^{\rho \sigma | \Sigma} +2 I_{\Lambda \Sigma}  F_{\mu \alpha}^{\Lambda} {F_{\nu}}^{\alpha|\Sigma} +g_{\mu \nu} V \,,
\end{align}
which, by calculating its trace and substituting, yields
\begin{align} \label{useful_einst}
R_{\mu \nu}  &\eq 2  g_{i \bar{\jmath}} \partial_{\mu}t^{i} \partial_{\nu}\bar{t}^{\bar{\jmath}} + 2  h_{uv} D_{\mu}q^u D_{\nu} q^v \nonumber \\
&\pheq + \frac12 I_{\Lambda \Sigma} g_{\mu \nu} F_{\rho \sigma}^{\Lambda} F^{\rho \sigma | \Sigma} -2 I_{\Lambda \Sigma}  F_{\mu \alpha}^{\Lambda} {F_{\nu}}^{\alpha|\Sigma} + g_{\mu \nu} V \,,
\end{align}
which is a useful expression to manipulate the on-shell action in section \ref{el_onshell}. The nonvanishing components of the Einstein's tensor read:
\begin{align}
E_{tt} &=  -\frac{h e^{-\beta} \left(r h'+h-1\right)}{r^2} \ , \nonumber \\
E_{rr} &= \frac{r h'-r h \beta'+h-1}{r^2 h} \ , \nonumber \\
E_{\theta \theta} = \frac{ E_{\phi \phi}}{\sin \theta^2} &= \frac{1}{4} r \left(2 r h''+h' \left(4-3 r \beta '\right)+h \left(-2 r \beta''+r (\beta')^2-2 \beta'\right)\right)  \ .
\end{align}
The $tt$ and $rr$ component of the Einstein's equations are equivalent to the following two equations \\
$tt$ + $rr$)
\begin{equation}
  \label{einsteintrr}  -\beta' =  \frac{1}{ \tau_1^2} \left[ (\partial_r \tau_1)^2 + (\partial_r b_1)^2 \right] +\frac{1}{2 \tau_2^2} \left[ (\partial_r \tau_2)^2 + (\partial_r b_2)^2 \right] +\frac{2}{\sigma^2} (\partial_r \sigma)^2 + \frac{24}{h^2} \sigma^4 \zeta^2 e^{\beta}
\end{equation}
$tt$)
\begin{align}  \label{einsteintt}
\frac{r h'+h-1}{r^2} &\eq -\frac{ \sigma^4 \left(4 b_1 b_2+2 b_1^2+6\right)^2+4\sigma^4 \left(2\tau_1^2 (b_1+b_2)^2+4 b_1^2\tau_2^2\right)}{4\tau_1^2\tau_2} \nonumber \\ 
&\pheq - \sigma^4 \left(\frac{\tau_1^2}{\tau_2}+2\tau_2\right) + 8 \sigma^2 \left(\frac{2}{\tau_1}+\frac{1}{\tau_2}\right) - 12 \sigma^4 \zeta^2 \frac{e^{\beta}}{h} \\
&\pheq - \frac{h}{4} \bigg[\frac{2}{\tau_1^2}[(\partial_r \tau_1)^2 + 2(\partial_r b_1)^2]+\frac{1}{\tau_2^2}[ (\partial_r \tau_2)^2 + (\partial_r b_2)^2] \bigg] - \frac{h}{\sigma^2} ( \partial_{r}  \sigma)^2  + \frac{V_{BH}}{r^4} \ , \nonumber
\end{align}
where we defined the so-called black hole potential $V_{BH}$ as
\begin{equation}
V_{BH} = -\frac12 \left(P^{\Lambda} \,, e_{\Lambda}(r) \right) \left( \begin{array}{cc} \im \mathcal{N}_{\Lambda \Sigma} + \re \mathcal{N}_{\Lambda \Gamma} \im \mathcal{N}^{\Gamma \Theta} \re \mathcal{N}_{\Theta \Sigma} & -\re \mathcal{N}_{\Lambda \Gamma} \im \mathcal{N}^{\Gamma \Sigma} \\ - \im \mathcal{N}^{\Lambda \Gamma} \re \mathcal{N}_{\Gamma \Sigma} & \im \mathcal{N}^{\Lambda \Sigma}
\end{array} \right) \left( \begin{array}{c} P^{\Sigma} \\ e_{\Sigma}(r) \end{array} \right)\,.
\end{equation}

We have checked that the system of equations \eqref{einsteintrr}-\eqref{einsteintt} implies the last two nonvanishing components of the Einstein's equations, as noticed in \cite{Gauntlett:2009bh} and \cite{Toldo:2012ec}. Therefore eqs. \eqref{maxw_e}, \eqref{scalar_tau1}, \eqref{scalar_tau2}, \eqref{scalar_b1}, \eqref{scalar_b2}, \eqref{scalar_sigma}, \eqref{einsteintrr}, \eqref{einsteintt} are the equations of motion we have to solve to find black hole configurations.